\documentclass[12pt,tightenlines,eqsecnum,floats,aps,amsmath,amssymb,nofootinbib,prd,shownopacs,floatfix]{revtex4-2}

\usepackage{mathptmx}        % selects Times Roman as basic font
\usepackage{amsmath}
\usepackage{amssymb}
\usepackage{color}
\usepackage{helvet}          % selects Helvetica as sans-serif font
\usepackage{courier}         % selects Courier as typewriter font
\usepackage{dirtree}
\usepackage{makeidx}        % allows index generation
\usepackage{graphicx}        % standard LaTeX graphics tool
\usepackage{subfig}

\usepackage[bottom]{footmisc}% places footnotes at page bottom

\usepackage{hyperref}        %for hyperlinks
\hypersetup{colorlinks=true,urlcolor=blue}
\usepackage[misc]{ifsym}
\usepackage{titlesec}

\titleformat{\section}
  {\normalfont\fontsize{15}{15}\bfseries}{\thesection}{1em}{}

\titleformat{\subsection}
  {\normalfont\fontsize{13}{15}\bfseries}{\thesubsection}{1em}{}

\usepackage[margin=1.3in]{geometry}

\usepackage{soul}

\makeindex             % used for the subject index
\usepackage{setspace}
\usepackage{graphicx}
\usepackage{enumerate}
\usepackage{color}
\usepackage{mathrsfs}
\usepackage{url}
\usepackage{caption}
\definecolor{mg}{rgb}{0.0, 0.5, 0.0}

\def\be{\nopagebreak[3]\begin{equation}}
\def\ee{\end{equation}}
\def\ba{\nopagebreak[3]\begin{eqnarray}}
\def\ea{\end{eqnarray}}
\newcommand{\f}{\frac}
\def\rmd{{\rm d}}
\def\lp{\ell_{\rm Pl}}
\def\d{{\rm d}}

\def\t{\tilde}

\def\db{\delta_b}
\def\dc{\delta_c}
\def\T{\mathcal{T}}
\def\g{\mathfrak{g}}
\def\P{\mathcal{P}}

\usepackage{mathrsfs}
\newcommand*{\scri}{\ensuremath{\mathscr{I}}}
\newcommand*{\scrip}{\ensuremath{\mathscr{I}^{+}}}
\newcommand*{\scrim}{\ensuremath{\mathscr{I}^{-}}}

\def\ulp{\underline{p}}

\def\DH{\emph{DH}\, }
\def\DHs{\emph{DHs}\,\,}
\def\TDH {\emph{T-DH}\, }
\def\ATDH {\emph{AT-DH}\, }

\begin{document}

%%%%%%%%%%%%%%%%%%%%%%%%%%%%%%%%%%%%%%%%%%%%%%%%%%%%%%%%%%%%%%%%%

\title{Regular black holes from Loop Quantum Gravity}
%\titlerunning{Black Holes from LQG}
\author{Abhay Ashtekar$^1$}
\email{ashtekar.gravity@gmail.com}
\author{Javier Olmedo$^2$}
\email{javolmedo@ugr.es}
\author{Parampreet Singh$^3$}
\email{psingh@lsu.edu}
\affiliation{$^{1}$ Physics Department, and, Institute for Gravitation \& the Cosmos, Penn State, University Park, PA 16802, USA \\
$^{2}$ Departamento de F\'isica Te\'orica y del Cosmos,
Universidad de Granada, Granada-18071, Spain \\
$^{3}$ Department of Physics and Astronomy, Louisiana State University, Baton Rouge, LA 70803, USA}

\begin{abstract}
There is rich literature on regular black holes from loop quantum gravity (LQG), where quantum geometry effects resolve the singularity, leading to a quantum extension of the classical space-time. As we will see, the mechanism that resolves the singularity can also trigger conceptually undesirable features that can be subtle and   are often uncovered only after a detailed examination. Therefore, the quantization scheme has to be chosen rather astutely. We illustrate the new physics that emerges first in the context of the eternal black hole represented by the Kruskal space-time in classical general relativity, then in dynamical situations involving gravitational collapse, and finally, during the Hawking evaporation process. The emphasis is on novel conceptual features associated with the causal structure, trapping and anti-trapping horizons and boundedness of invariants associated with curvature and matter.
This Chapter is not intended to be an exhaustive account of all LQG results on non-singular black holes. Rather, we have selected a few main-stream thrusts to anchor the discussion, and provided references where further details as well as discussions of related developments can be found. In the spirit of this Volume, the goal is to  present a bird's eye view that is accessible to a broad audience.\footnote{Invited Chapter for the book Regular Black Holes: Towards a New Paradigm of Gravitational Collapse, Ed. C. Bambi, Springer Singapore (2023)}
\end{abstract}

\maketitle
%%%%%%%%%%%%%%%%%%%%%%%%%%%%%%%%%%%%%%%%%%%%%%%%%%%%%%%%%

\section{Introduction}
\label{s1}

%Use the template \emph{chapter.tex} together with the Springer document class svmult to style the various elements of your chapter content in the Springer layout. However, authors should not worry too much about the exact style, as this can be fixed by the Editorial Office in the preparation of the proof.

There is general agreement in the gravity community that black hole singularities of classical general relativity (GR) offer excellent opportunities to probe physics beyond Einstein. However, as of now, there is no consensus on the fate of black hole singularities in full quantum gravity. Indeed, there is an ongoing debate even on a central question in the subject: Will singularities of classical GR be naturally resolved in full quantum gravity, or will they persist? As the very name of this Volume suggests, in many circles an affirmative answer is taken to be a necessary condition for the viability of a proposed quantum gravity theory. But this is not an universally accepted  viewpoint. For example, it has been argued that taming of black hole singularities in asymptotically anti-deSitter space-times would violate a ``No Transmission Principle" motivated by the  AdS/CFT correspondence \cite{eh}. More generally, discussions of the black hole evaporation process are often based on the assumption that there is a singularity also in quantum gravity. These expectations are based on the Penrose diagram of an evaporating black hole that Hawking drew over 40 years ago \cite{swh}, where the singularity persists as part of the future boundary of space-time even after the black hole has completely disappeared (see Fig. \ref{fig:trad}). However, this feature of the diagram was not arrived at from a calculation, and indeed such a calculation is not available even today. Furthermore, some forty years later Hawking himself changed his mind: A new Penrose diagram was proposed to represent an evaporating black hole in which there is no singularity (see Fig. 2 of \cite{hps}). Nonetheless, interestingly, Hawking's first paradigm continues to feature prominently in discussions on the issue of information loss: see, e.g., Ref. \cite{uw} where the persistence of this singularity leads to a non-unitary evolution from $\scrim$ to $\scrip$, and Refs. \cite{apms,giddings,marolf} where proposals are made on how unitarity could be rescued in spite of this singularity, thereby preventing information loss.

Loop quantum gravity (LQG) provides a systematic avenue to investigate the fate of singularities of classical GR because it is based on \emph{quantum} Riemannian geometry. Consequently, new physics arises in the Planck regime where the continuum space-time of classical GR becomes inadequate (see, e.g., \cite{apbook}). Implications of this new physics have been analyzed \emph{in detail} in the commonly used cosmological models. Non-perturbative quantum corrections to Einstein's equations imply that, once a curvature invariant approaches the Planck scale, quantum geometry modifications of Einstein dynamics introduce strong `repulsive corrections' that dilute that invariant, preventing a blow-up (see, e.g., \cite{asrev,ps,iapsrev}). Thus, the big-bang/big-crunch singularity is replaced by a quantum bounce in loop quantum cosmology (LQC). Once the curvature drops to about $\sim\! 10^{-4}$ Planck scale, quantum corrections can be neglected and classical GR becomes a good approximation.

A natural question then is whether the same phenomenon occurs at the black hole singularities. Results to date provide considerable evidence that it does. However, technically, the situation is more complicated than that in cosmological models for two reasons. First, even in the Schwarzschild solution, although space-time is homogeneous in the vicinity of the singularity, it is \emph{not} isotropic. Second, the nature of the blow up of curvature is different from that in the commonly used cosmological models: As Penrose has emphasized, while the Weyl curvature vanishes identically at the big-bang in homogeneous isotropic cosmologies, it diverges at the Schwarzschild singularity. As a result, although the singularity is resolved in all LQG investigations, as of now, results in the black hole sector are not as strong as they are in LQC. Nonetheless, a large number of investigations, carried out since 2004, have provided conceptual insights as well as detailed technical results on the nature of the resolution of the Schwarzschild singularity.
Our goal is to convey an overall picture at a technical level that is accessible to beginning researchers, emphasizing conceptual issues, novel elements, and problems that remain. We also provide references where details can be found. Also for convenience of non-experts, throughout the Chapter, we pause to summarize the main points after each technical discussion and also at the end of subsections.

In Sections \ref{s2} and \ref{s3} we focus on the quantum extension of the Kruskal space-time. Because the static Killing field is space-like in the `interior' region --bounded by the singularity in the future and the horizon in the past-- the space-time metric is spatially homogeneous (but not isotropic). As is well-known, this portion of Kruskal space-time is isometric with the vacuum Kantowski-Sachs cosmological model. Therefore techniques from LQC have been used to analyze the fate of the Schwarzschild singularity in a number of investigations within LQG %\cite{ab} - \cite{ao}.
(See, e.g.,\cite{ab,lm,cgp,gp,ck,bv,dc,cs,oss,cctr,yks,js,dc2,bkd,gop,djs,aoslett,aos,ao,nbetal,maetal:2020,gop:2021,maetal:2022,menaetal0,mena-quis,menaetal,han}).
While some of these analyses present us with the equations that dictate the evolution of the \emph{quantum state} of the system, the detailed results are based on the so-called `effective equations' whose goal is to incorporate the leading order quantum corrections to the classical geometry in sharply peaked quantum states.\footnote{For the conceptual framework underlying effective equations see, e.g., Section V of \cite{asrev}. Note that the term `effective equations' has a very different connotation here than in standard quantum field theory. This has caused occasional confusion in the literature. In LQG one \emph{does not integrate out} `high energy modes'; Planck scale effects are retained. In LQC, for example, there are states that remain sharply peaked even in the Planck regime and the effective equations capture the evolution of the peak of the quantum wave function in these states, ignoring the fluctuations.}
At a conceptual level, all these investigations follow the same strategy. However, the technical implementation of this procedure differs, leading to different effective geometries in the interior region. Nonetheless, in all these cases, the singularity is resolved due to quantum corrections. We will discuss the strategy and compare and contrast various results in Section \ref{s2}. Singularity resolution in the Kruskal space-time provides several sharp results on the causal structure of its quantum extension. In particular, the singularity is replaced by a `transition surface' to the immediate past of which we have a trapped region and to the immediate future, an anti-trapped region. This geometry is sometimes referred to as depicting `a black hole to white hole transition'. We will avoid this terminology because it has other connotations that are not realized. In particular, the terms `black hole' and `white hole' normally go hand in hand with singularities and event horizons. In LQG, singularities are absent and, in dynamical situations, there are also no event horizons either.

In Section \ref{s3} we consider the Schwarzschild exterior, i.e. the region bounded by the horizon and $\scri^\pm$. Space-time is again foliated by homogeneous 3-dimensional surfaces but they are now time-like rather than space-like. We discuss a possible extension of the `interior' geometry to this exterior region, following \cite{hlkn,aoslett,aos}. This extension has several attractive properties \cite{ao}, but it also has some puzzling features: while the quantum corrected metric is again asymptotically flat in a precise sense (that suffices to define the ADM mass, for example), the approach to the flat metric is weaker than the one generally used in the physics literature. There \emph{are} alternate  proposals to arrive at effective metrics with the standard asymptotic behavior (see, e.g., \cite{gop:2021,han}) but a definitive picture is yet to emerge.

Now, the Kruskal space-time itself is an idealization since it represents an `eternal black hole'; black holes encountered in nature are formed \emph{dynamically}, e.g., via a gravitational collapse, or compact binary mergers. Nonetheless, one would expect the qualitative features of the causal structure that arises from taming of the singularity due to quantum effects would be robust. In Section \ref{s4} we discuss models of dynamical situations that have been analyzed within LQG and summarize the current status, focusing on the Lema\^{i}tre-Tolman-Bondi type models of collapse and critical phenomena discovered by Choptuik. In Section \ref{s5} we turn to the issue of black hole evaporation and `information loss'. The LQG discussion of these issues is characterized by two key features \cite{aamb}. First, as discussed above, in contrast to the Penrose diagram in Hawking's seminal paper \cite{swh}, there is no singularity in the space-time interior which can serve as a `sink of information'.  Second, as the LQG Penrose diagram of Fig.~\ref{fig:LQG} shows, there is no event horizon: what forms and evaporates is a dynamical horizon \cite{ak2,akrev,boothrev}. Much of the discussion in the literature assumes that there is an event horizon which serves as a boundary of an `interior' region from which no causal signal can ever be sent to the asymptotic region. One is then led one to either conclude that information is lost, or, to introduce `exotic' ideas such as quantum Xerox machines, firewalls and fast scramblers to restore unitarity. As we discuss, there is a more direct pathway to unitarity once it is realized that there is no event horizon. However, as in every other approach, important issues remain: the precise nature quantum radiation at the final stages of the evaporation process require full LQG and this analysis has only begun. We summarize the current status in Section \ref{s5}. In Section \ref{s6} we collect the key features of regular black holes in LQG compare and contrast the regular LQG black holes with this in other approaches.

Our conventions are the following. Space-time metric $g_{ab}$ has signature -,+,+,+ and the curvature tensors are defined by $R_{abc}{}^d k_d = 2 \nabla_{[a} \nabla_{b]} k_c;\,\, R_{ac} = R_{abc}{}^b$;\,\, and $R = g^{ab} R_{ab}$. By macroscopic black holes we mean those for which $GM =:m \gg \lp$.

\section{The Schwarzschild interior}
\label{s2}

\begin{figure}[b]
%\sidecaption[t]
\includegraphics[width=0.62\textwidth]{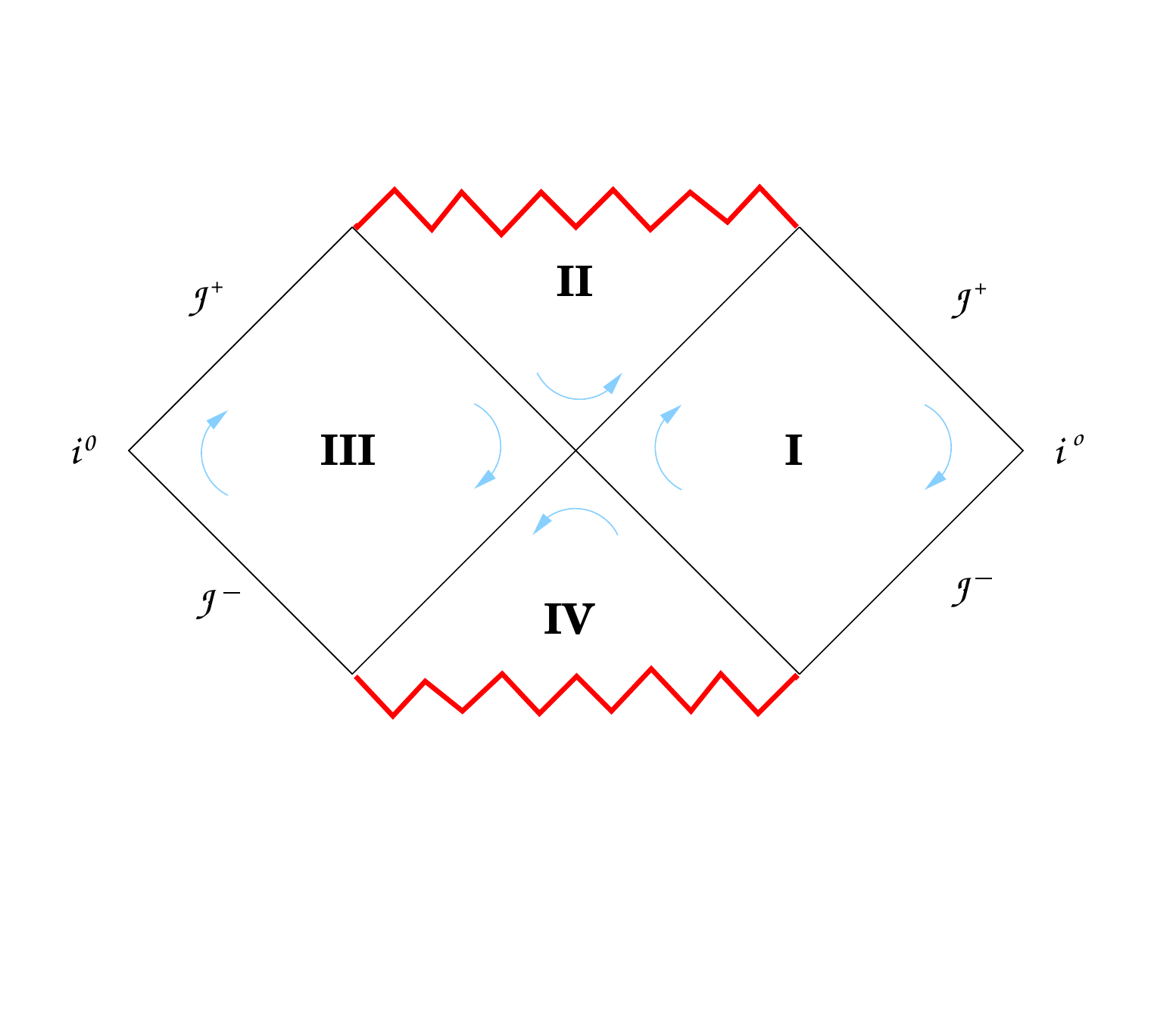}
%\caption{\vskip0.7cm \footnotesize{
\caption{ \footnotesize{The Penrose diagram of the Kruskal space-time. In this section we discuss the quantum extension of part II, bounded to the past by future horizons and the future by the singularity. The quantum corrected effective geometry of region I is discussed in Section \ref{s2.2}.}}
\label{fig:kruskal}
\end{figure}

Denote by $(M, g_{ab})$ the Kruskal extension of the Schwarzschild metric (see Fig. \ref{fig:kruskal}) and by $(M_{\rm II},\, g_{ab})$ the quadrant of this space-time that represents the (open) `interior region' II, bounded by the black hole singularity and future horizons. This region is foliated by the $r_{\rm sch}={\rm const}$ space-like manifolds, with topology $\mathbb{S}^2\times\mathbb{R}^2$. Each leaf admits 3 rotational Killing fields tangential to its 2-dimensional spherical cross sections that are mapped to one another by the translational Killing field.
%
%\nopagebreak[3]\begin{figure}[b] %\bfig
%\vskip0.3cm\includegraphics[width=2.5in,height=2.2in]{classical-kruskal.pdf} %\hskip1cm
\nopagebreak[3]
Consequently, $(M_{\rm II},\, g_{ab})$ is spatially homogeneous, but not isotropic; it is isometric to the (vacuum) Kantowski-Sachs cosmological model. Therefore LQG approaches use the procedure from homogeneous cosmologies. Now, while the big-bang and big-crunch singularities persist in the Wheeler-DeWitt (WDW) theory based on metric variables,  they are naturally resolved in LQC because of the quantum geometry resulting from the use of  connection variables (see, e.g., \cite{asrev}). For the Schwarzschild interior, then, LQG investigations also begin with a 3+1 decomposition of Einstein's equations using connection variables. In the classical theory, components of the curvature tensor that features in these equations can be obtained by first evaluating holonomies of the gravitational connections around suitable closed loops (called  plaquettes) and then taking the limit as the area enclosed by these plackets tends to zero. In LQG, the corresponding quantum operator is obtained by shrinking these plaquettes till the area they enclose reaches the smallest non-zero eigenvalue of the area operator. This eigenvalue is called \emph{the area gap} and denoted by $\Delta$. As a consequence, information about quantum geometry gets encoded in the dynamical equations. Observables such as curvature scalars can acquire \emph{finite} upper bounds on entire dynamical trajectories, whence the singularity is resolved. $\Delta$ appears in the denominator of the expressions of these upper bounds; classical singularities emerge as $\Delta \to 0$.

For black holes, while operator equations have been written down \cite{ab,lm,cgp,gp,maetal:2022,menaetal}, detailed investigations of the singularity resolution and ensuing quantum corrected geometry have been obtained using `effective equations' discussed in Section \ref{s1}. Solutions to effective equations show that the central singularity is resolved due to quantum corrections. However, different investigations within LQG have made different choices to arrive at the quantum corrected curvature operators. Intuitively these choices represent quantization ambiguities that then affect detailed predictions. For brevity, in Sections \ref{s2.1} and \ref{s2.2} we will present the general framework and results following a recent approach that is free of limitations of the earlier investigations and in Section \ref{s2.3} we will briefly compare and contrast other approaches. Due to space limitation, by and large we will only include motivations behind various constructions and summarize the final results. For detailed derivations and other details, see in particular \cite{aos,cs,oss,ab,lm}.

\subsection{The framework}
\label{s2.1}

In connection-dynamics, the initial data for space-time geometry consists of an SU(2)-valued  connection $A_a^i$ and its conjugate `electric field' $E^a_i$ as in Yang-Mills theory. In the final solutions to Einstein's equations, $A_a^i$ has the interpretation of the gravitational connection that parallel transports SU(2) spinors, and $E^a_i$, represent the ortho-normal spatial triads (with density weight 1). Because of spatial homogeneity of the model, various spatial integrals in the Hamiltonian framework have a trivial divergence. Therefore, one introduces an `infrared cut-off'. Thus one truncates the homogeneous slices to be finite (rather than infinite) cylinders, with coordinates $(\theta,\phi, x)$ with $x\in (0,\, L_{\circ})$ (rather than $x \in (0,\,\infty)$). One has to make sure, of course, that none of the final results depend on $L_\circ$. One can solve the `kinematical' constraint equations and use gauge-fixing to cast the basic variables in the form
\ba \label{AE}
A^i_a \, \tau_i \, \d x^a \, &=& \, c/L_\circ \, \tau_3 \, \d x + b\,( \tau_2 \d \theta
- \tau_1 \sin \theta \, \d \phi) + \tau_3 \cos \theta \, \d \phi,\nonumber\\
E^a_i \, \tau^i \partial_a \, &=&  \, p_c \, \tau_3 \, \sin \theta
\, \partial_x + (p_b/L_\circ)\, \tau_2 \, \sin \theta \,
\partial_\theta - (p_b/L_\circ)\,\, \tau_1 \,  \partial_\phi
\ea
where $\tau_i$ are SU(2) generators related to Pauli spin matrices $\sigma_{i}$ via  $\tau_i = - i \sigma_i/2$.\,
Real valued connection components ${b},{c}$ and the triad components ${p}_{b}, {p}_{c}$ are functions only of time and serve as conjugate coordinates on the 4-dimensional phase space. It is convenient to choose an orientation of the triads so that $b,\,c,\,p_c$\, are positive and $p_b$ is negative. It follows from (\ref{AE}) that physical quantities can only depend on $b,\, (p_b/L_\circ),\, (c/L_\circ),\, p_c$.\, Given a time coordinate $\tau$ that labels the spatially homogeneous surfaces and the corresponding lapse $N_{\tau}$, in region II the space-time metric has the form
\be\label{metric}
g_{ab} \d x^{a} \d x^{b} \equiv \d s^2 = - N_{\tau}^2 \d \tau^2 + \f{p_b^2}{p_c L_\circ^2} \d x^2 + p_c (\d \theta^2 + \sin^2\theta \d \phi^2) .
\ee
At the horizon, $b, p_b$ vanish and the translation Killing field\, $X = \partial/\partial x$\, becomes null. When  $p_c$ vanishes, the radius of the metric 2-spheres shrinks to zero, making the curvature scalars diverge there. This is Schwarzschild singularity.

It turns out that Einstein's equations that govern the dynamics of the basic variables simplify significantly if one uses the lapse  $N_{\rm cl}= (\gamma\, \sqrt{p_c})/b$ (which is different from the standard lapse in the Schwarzschild coordinates.) The $\gamma$ in this expression is the dimensionless Barbero-Immirzi parameter of LQG. It is analogous to the $\theta$-parameter of QCD in that it represents a quantization ambiguity: classical physics is insensitive to the precise value of $\gamma$; we only need $\gamma >0$. In terms of the corresponding time-coordinate $T_{\rm cl}$, the dynamical trajectories are given by:
\be \label{conf}
b(T_{\rm{cl}})= \gamma\, \left(e^{-T_{\rm{cl}}}-1\right)^{1/2} \quad {\rm and} \quad
p_b(T_{\rm{cl}})=p_b^{(\circ)}\, e^{T_{\rm{cl}}}\,\big(e^{-T_{\rm{cl}}} - 1\big)^{1/2},
\ee
and
\be \label{momenta}
 c(T_{\rm{cl}}) =   \,c_{(\circ)}\,e^{-2 T_{\rm{cl}}} \quad {\rm and} \quad p_c(T_{\rm{cl}}) = p_c^{(\circ)}\,e^{2 T_{\rm{cl}}} ~.
\ee
Here\, $c_{(\circ)},\, p_b^{(\circ)},\,p_c^{(\circ)}$ are integration constants. Comparison with the standard form of the Schwarzschild solution yields $p_c^{(\circ)} = 4 m^2$,\, $p_b^{(\circ)}/L_\circ = -2m$,\, and \, $c_{(\circ)}/L_\circ = \gamma/4m$, where $m$ is related to the mass of the Schwarzschild solution via $m=GM$. At the horizon $T_{\rm cl} =0$ and at the singularity $T_{\rm cl} = -\infty$.

The dynamical variables are subject to the Hamiltonian constraint
\be\label{Hcl}
H_{\mathrm{cl}}[N_{\rm{cl}}] \equiv - \f{1}{2 G \gamma}\Big(\big(b + \f{\gamma^2}{b}\big) p_b\,+\, 2 c \, p_c   \Big) =0.
\ee
It is easy to verify that the terms in the $b$ and $c$ sectors on the right side of (\ref{Hcl}) are separately conserved in time, and equal $-m$ and $m$ respectively on solutions. Therefore, if the constraint (\ref{Hcl}) is satisfied at one instant $T_{\rm cl}$, then it holds for all $T \in (-\infty, 0)$.

As explained above, in the passage to quantum theory the spatial curvature is expressed using the holonomoly of the gravitational connection $A_a^i$ around appropriately chosen plaquettes that enclose the minimum non-zero area, $\Delta = 4\sqrt{3}\,\pi\gamma\lp^2$. (Thus, while classical physics is insensitive of the value of the Barbero-Immirzi parameter $\gamma$, quantum physics is not. Its value is generally taken to be $\gamma= 0.2375$ via black hole entropy calculation.) As a consequence, the effective equations that capture the leading quantum corrections inherit new \emph{`quantum parameters'}, denoted by $\delta_b$ and $\delta_c$, that refer to edge lengths of these plackets, and go to zero in the classical limit, $\lp \to 0$ (or, $\Delta \to 0$, keeping $\gamma$ fixed). Different choices of these quantum parameters represent quantization ambiguities mentioned above. In this section we will use a strategy \cite{aoslett,aos,ao} that is free of the physically undesirable features encountered in other approaches (discussed in Section \ref{s2.3}).

A key idea behind this strategy is to use $\delta_b$ and $\delta_c$ that are `Dirac observables' i.e. phase space functions that are \emph{constant along dynamical trajectories}.\footnote{Because the spatial curvature features on the right side of Einstein's evolution equations, the quantum corrected version of the classical dynamical trajectories (\ref{conf}) and (\ref{momenta}) along which $\delta_b$ and $\delta_c$ are to remain constant themselves feature $\delta_b$ and $\delta_c$ (see (\ref{eq:c}), (\ref{eq:b}) and (\ref{eq:pb})). Therefore the issue of finding $\delta_b$ and $\delta_c$ that are Dirac observables is rather subtle conceptually and quite intricate technically. These subtleties has led to some concerns \cite{nbetal}. This issue is analyzed in detail \cite{mena-quis,menaetal0,menaetal,Ongole:2022rqi}. Consistency of the final results directly follows from the effective equations (\ref{eomb}) - (\ref{H_eff}).}
Let us restrict ourselves to such $\delta_b,\,\delta_c$ from now on. Then, again, the evolution equations simplify if we include the appropriate quantum corrections in the choice of the lapse, defining it as $N:= (\gamma \,\sqrt{p_c})\,{\delta_b}/{\sin(\db b)}$. (Note that as the area gap $\Delta$ goes to zero, so does $\delta_b$ and $N$ reduces to $N_{\rm cl}$.)\, Denote by $T$ the corresponding time parameter and by `dot' the derivative with respect to $T$. Then, as in the classical theory, the effective evolution equations $b$ and the $c$ sectors separate:
\be \label{eomb}
\dot b = - \f{1}{2} \left(\f{\sin(\db b)}{\db} +\f{\gamma^2  \db}{\sin(\db b)}\right), \qquad
\dot p_b = \f{p_{b}}{2} \, \cos(\db b) \left(1 - \f{\gamma^2  \db^2}{\sin^2(\db b)}\right) ,
\ee
and
\be \label{eomc}
 \dot c = - 2 \, \f{\sin(\dc c)}{\dc}, \qquad \dot p_c = 2 \, p_c \, \cos(\dc c) .
\ee
But, again as in the classical theory, the two sectors are linked by the (now, effective) Hamiltonian constraint:
\be \label{H_eff}
H_{\mathrm{eff}}[N] \equiv  - \f{1}{2 G \gamma} \Big[ \Big(\f{\sin (\delta_b b)}{\delta_b} + \frac{\gamma^2 \delta_b}{\sin(\delta_b b)} \Big) \, p_b + 2 \f{\sin (\delta_c c)}{\delta_c}  \, p_c  \Big] =0 .  \ee
A direct calculation shows that the constraint (\ref{H_eff}) is preserved in time.

To summarize, conditions $\dot\db =0,\,\dot\dc=0$, the evolution equations (\ref{eomb}), (\ref{eomc}) and the constraint equation (\ref{H_eff}) constitute a set of consistent equations that generalize the classical constraint and evolution equations. A notable difference from the classical theory arises because in LQG there is a well-defined operator in the quantum theory corresponding only to the holonomy defined by the gravitational connection $A_a^i$, rather than $A_a^i$ itself. As a consequence, only trigonometric functions of $\db\,b$ and $\dc\,c$ appear. Hence the domain of these variables is compactified (just as in LQC \cite{aps}): they take values in the open interval $(0,\, \pi)$. The momenta $p_b,\,p_c$, by contrast, continue to assume values $p_b <0$ and $p_c >0$ as in the classical theory.

To solve the evolution equations, it is convenient to first obtain solutions $c(T), p_c(T)$ and $b(T)$. Now, in the $c$ sector, equations of motion (\ref{eomc}) immediately imply that $m_c =  {(\sin(\delta_c c)\, p_c)}/({\gamma L_\circ \delta_c})$ is a constant of motion. This fact simplifies the form of the solutions. One obtains
\ba \label{eq:c}
\tan \Big(\f{\delta_{c}\, c(T)}{2} \Big) &=&   \f{\gamma L_o \dc}{8 m_c} e^{-2 T},\qquad
p_c(T) = 4 m_c^2 \Big(e^{2 T} + \f{\gamma^2 L_o^2 \dc^2}{64 m_c^2} e^{-2 T}\Big),\quad \\
\label{eq:b}
\cos \big(\delta_{b}\,b(T)\big) &=& b_o \tanh\left(\f{1}{2}\Big(b_o T + 2 \tanh^{-1}\big(\frac{1}{b_o}\big)\Big)\right)\,, \ea
where there constant $b_o$ is given by $b_o = (1 + \gamma^2 {\delta_b}^2)^{1/2}$.\, One then uses the Hamiltonian constraint to determine $p_b(T)$:
\be \label{eq:pb}
p_b(T) = - 2 \f{\sin (\dc\, c(T))}{\dc} \f{\sin (\db\, b(T))}{\db} \f{p_c(T)}{\f{\sin^2(\db\, b(T))}{\db^2} + \gamma^2} .
\ee
Eqs (\ref{eq:c}) - (\ref{eq:b}) provide the dynamical trajectories of the effective theory. It is easy to verify that in the limit\, $\db \to 0,\, \dc \to 0$,\, one recovers the classical trajectories. To summarize, the quantum corrected, effective trajectories are given by (\ref{eq:c}) - (\ref{eq:pb}) for any choice of constants of motion $\delta_b,\,\delta_c$. Since these equations only involve the combinations\, $b,\,(p_b/L_\circ),\, \delta_b;\, (c/L_\circ), p_c,$\, and\, $L_\circ\delta_c$,\, the metric (\ref{metric}) and all physical results are insensitive to the choice of the infrared cut-off $L_\circ$.

So far $\db,\,\dc$ could be any quantum parameters satisfying  $\dot\db=0$ and $\dot\dc=0$. The following considerations provide a natural avenue to determine them. Recall that on classical solutions, the $c$ part of $H_{\mathrm{cl}}[N_{\rm{cl}}]$ equals $m$,\, and the $b$ part equals $-m$. Therefore in the effective theory, one is led to set
\be \f{1}{2\gamma}\Big(\f{\sin (\delta_b b)}{\delta_b} + \frac{\gamma^2 \delta_b}{\sin(\delta_b b)} \Big) \,\f{p_b}{L_\circ}\, = -m_b \quad {\rm and}\quad  \f{\sin(\delta_c c)}{\gamma L_\circ \delta_c} = m_c\, .    \ee
Equations of motion (\ref{eomb}) and (\ref{eomc}) imply that both $m_b$ and $m_c$ are constants of motion and
the effective Hamiltonian constraint reads $m_b = m_c$. On solutions, we will drop the suffix and set $m_b=m_c =m$.
The fact that $m_b$ and $m_c$ are constants of motion suggests a natural strategy to restrict the form of $\delta_b,\delta_c$: Require that $\db$ be a function only of $m_b$, and $\dc$ be a function only of $m_c$. To constrain the functional form requires additional input, summarized in Section \ref{s2.2}. Here we only note that the final answer has a rather simple form for large black holes (i.e. for solutions for which $m \gg \lp$): $\db$ and $\dc$ are extremely well-approximated by
\be\label{db-dc}
\db=\Big(\frac{\sqrt{\Delta}}{\sqrt{2\pi}\gamma^2 m_b}\Big)^{1/3}, \qquad {\rm and}\qquad
L_{o}\dc=\frac{1}{2} \Big(\frac{\gamma\Delta^2}{4\pi^2 m_c}\Big)^{1/3}.
\ee
(Recall that physical results can only depend on the combination $L_{o} \delta_{c}$.)

To summarize, the effective metric in the interior region is given by (\ref{metric2}), where $c,p_c$ are given by (\ref{eq:c}), $b,p_b$ by (\ref{eq:b}),\, (\ref{eq:pb}),\, and\, $\delta_b,\,\delta_c$\, by\, (\ref{db-dc}). By inspection we see that as the area gap $\Delta$ goes to zero, $\db$ and $\dc$ both go to zero and the effective theory reduces to the classical GR.

\subsection{Singularity Resolution, Causal Structure and Curvature Bounds}
\label{s2.2}

Let us explore properties of the space-time metric
\be\label{metric2}
g_{ab} \d x^{a} \d x^{b} \equiv \d s^2 = - N^2 \d T^2 + \f{p_b^2}{p_c L_\circ^2} \d x^2 + p_c (\d \theta^2 + \sin^2\theta \d \phi^2) .
\ee
of the effective theory. The past boundary of the open region under consideration is again given by\, $b=0,\, p_b=0$\, which occurs at $T=0$ on every dynamical trajectory. The translational Killing vector $X^a$ becomes null at these points; thus as in the classical theory, this boundary represents the horizon. In the classical theory, the singularity is characterized by the vanishing of the radius of the metric 2-spheres, i.e., of $p_c$.\, In the effective theory, however, $p_c$ has a \emph{non-zero} minimum, $p_c^{\rm min} = \f{1}{2} \gamma (L_\circ \delta_c) m$ which occurs at $T = \frac{1}{2}\,\ln (\gamma L_\circ \delta_c)/8m$. Note that this minimum radius is of Planck scale but depends on the mass of the initial black hole:  $r_{\rm min} \sim (m\lp^{2})^{1/3}$. This is the surface that replaces the classical singularity and  the space-time metric  (\ref{metric2}) can be smoothly extended across this 3-manifold.

\begin{figure}[b] %\bfig
%\sidecaption[t]
\includegraphics[width=0.5\textwidth]{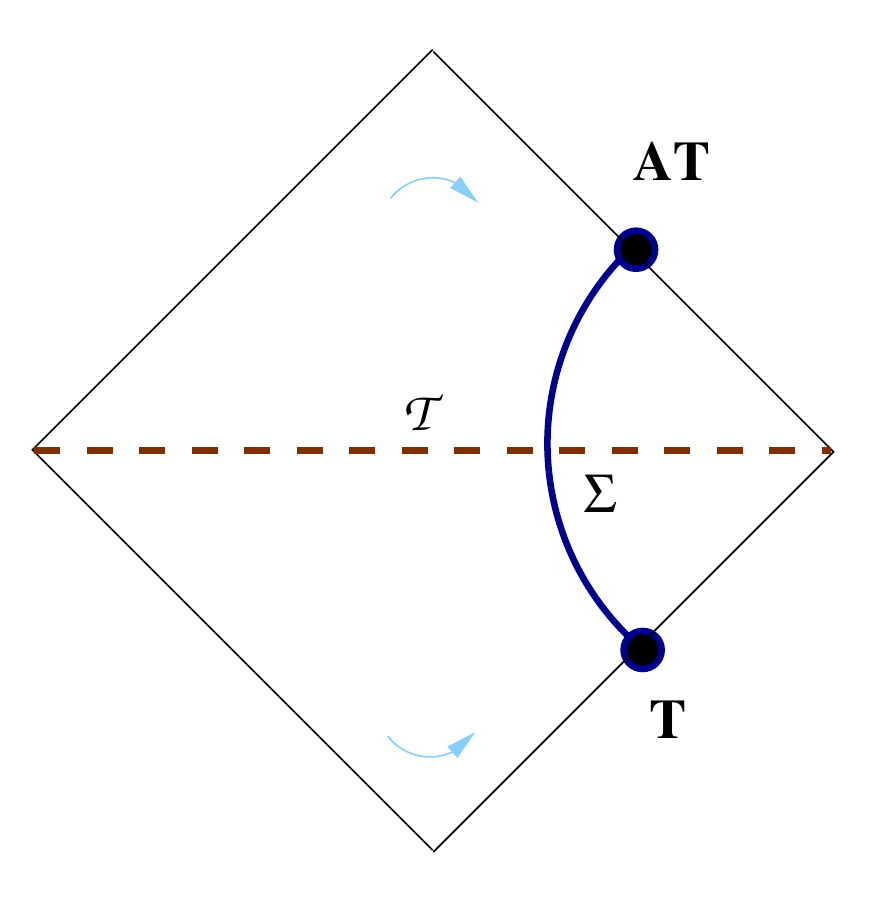}
%
%\caption{\footnotesize{
\caption{\footnotesize{Quantum extension of region II of Fig. \ref{fig:kruskal} in the effective theory. The singularity is replaced by the transition surface $\mathcal{T}$. It separates the trapped %region that lies to its past and the
and anti-trapped regions. The past boundary \textbf{T} is a null (black hole type) trapping horizon and the future boundary \textbf{AT} is a null (white-hole type) anti-trapping horizon. The time-like 3-manifold $\Sigma$ joining 2-spheres lying on the two horizons is used in Eq. (\ref{balance}). }}
\label{fig:diamond}
\end{figure}

One can explore the causal structure around this surface by calculating the expansions $\Theta_\pm$ of the two null normals to the metric 2-spheres. To the past of this surface one finds that both expansions are negative. Thus this is a \emph{trapped region} just as the entire region II is in the classical theory. Interestingly, \emph{both} null-expansions vanish on this surface. This is a novel situation that is not encountered in classical GR. Since the metric is smooth across this surface, space-time is well-defined across it and one can analyze the two expansions to the future of this surface. They are both positive, so the region to the future is \emph{anti-trapped}. Thus in the quantum-extended effective space-time,  the surface neatly separates a trapped region and an anti-trapped region. Therefore it is called a \emph{transition surface}, denoted by $\T$. It is analogous to the `bounce surface' in LQC (that replaces the big-bang), to the past of which the expansion of the universe is negative and to the future of which it is positive. However, now the term `expansion' refers to changes in the areas of metric 2-spheres along its two null normals. How far into the future is the space-time extended by this procedure? The metric is well defined in the open region bounded by the  surface $T= -(4/b_o)\, tanh^{-1}\,(1/b_o)$ where $(\delta_b\, b)=\pi$ and $p_b =0$. The Killing field $X^a$ is again null on the boundary so it again represents a horizon that bounds the anti-trapped region to the future. In summary, effective dynamics extends the open region II (of Fig. \ref{fig:kruskal}) to the diamond shaped open region (shown in Fig. \ref{fig:diamond}) bounded by Killing horizons. The region is separated by a transition surface $\T$, to the past of which one has a trapped region and to the future of which, an anti-trapped region. This extension is often referred to as the black hole to white hole transition.
\nopagebreak[3]

In LQC, space-time curvature attains the maximum value on the bounce surface and, furthermore, this upper bound is universal. Does the quantum corrected geometry exhibit the same feature at transition surface $\T$?\, The answer is in the affirmative. One has:
\ba \label{bounds}
R^{2}\mid_{\T}\,\,&\approx&\,\,\frac{256\pi^{2}}{\gamma^4\Delta^{2}}+ \ldots \qquad R_{ab}R^{ab}\mid_{\T}\,\,=\,\,\frac{256\pi^2}{\gamma^4\Delta^2}+ \ldots \\
C_{abcd}C^{abcd}\mid_{\T}\,\,&=&\,\, \frac{1024\pi^2}{3\gamma^4\Delta^2} + \ldots\qquad R_{abcd}R^{abcd}  =\,\, \frac{768\pi^2}{\gamma^4\Delta^2}+\ldots
\ea
where all the correction terms $\ldots$ have the same form\,\, ${\cal O}\big(({\Delta}/{m^2}))^{1/3} \,\ln ({m^2}/{\Delta})\big)$.\,\, Recall, first, that the classical limit corresponds to $\Delta \to 0$ (keeping $\gamma >0$.) Hence in this limit all invariants diverge and $\T$ is replaced by the singularity. Secondly, since leading terms are \emph{mass independent}, the upper bounds are universal. (The numerical coefficients vary simply because the invariants refer to distinct parts of the total curvature.) Third, as one moves away from $\T$, these curvature scalars rapidly approach their classical values even for very small black holes. Thus quantum corrections to space-time geometry are very small away from the transition surface. For instance, while the horizon radius of the effective solution is always larger than that of its classical counterpart, even for $m=10^4 \lp$, the relative difference is $\sim 10^{-15}$ and for a solar mass black hole, it is $\sim 10^{-115}$! Finally one can ask for the relation between the radius $r_{{}_{\rm T}}$ of the trapping horizon that constitutes the past boundary of the diamond, and the radius  $r_{\rm AT}$ of the anti-trapping horizon that constitutes the future boundary. Are they approximately the same? The answer is in the affirmative for macroscopic black holes, even though the `bounce' is not exactly symmetric. For a stellar mass black hole for example, $r_{{}_{\rm T}} = 3$km and $r_{\rm AT} = (3 + {O}(10^{-25}))$km.\, As we will see  in Section \ref{s2.3}, these consequences of effective dynamics are non-trivial: it is surprisingly difficult to achieve the singularity resolution without, at the same time, triggering unintended large effects away from the singularity.

Next, note that while the Ricci tensor vanishes identically in classical solutions, it is non-zero in the effective solutions. One can simply set $8\pi G_{\rm N}\, T_{ab}^{\rm eff} :=R_{ab} - \f{1}{2} R g_{ab}$ and interpret $T_{ab}^{\rm eff}$ as the effective stress-energy tensor of the quantum corrected space-time. As one would expect from the above  discussion, for macroscopic black holes these quantum corrections are negligible away from $\T$.\, However, they become large and dominant in the immediate vicinity of $\T$. As one could have anticipated, although it is finite everywhere, the energy density defined by $T_{ab}^{\rm eff}$ becomes large and negative in this region thereby violating the energy conditions, as it must for the singularity resolution to occur.

Interestingly, this fact creates an apparent tension with considerations involving the Komar mass $M_{\rm K}$. Recall that, in the classical theory,  $M_{\rm K}$ defined by the translational Killing field $X^a$ is given by (half the) horizon radius. As we saw, for macroscopic black holes the radii $r_{{}_{\rm T}}$ and  $r_{{}_{\rm AT}}$ are essentially the same. But the difference between the Komar mass evaluated at the anti-trapping horizon and the trapping horizon is given by the integral over a 3-manifold $\Sigma$ joining a cross-section of the trapping horizon with a cross-section of the anti-trapping horizon (see Fig.~\ref{fig:diamond}),
\be \label{balance} M_{\rm K}^{\rm AT} - M_{\rm K}^{\,\rm T}\, =\, 2\,\int_{\Sigma}\! \big(T_{ab}^{\rm eff}\, -\, \f{1}{2} T^{\rm eff}\,\, g_{ab}^{\rm eff}\big)\, X^a \rmd \Sigma^b \, ,\ee
and for macroscopic black holes the integrand of the right is \emph{large and negative} near $\T$ (because it represents the effective energy density). How can the two Komar masses be the same, then?  It turns out that the integrand of (\ref{balance}) is indeed large and negative for macroscopic black holes, but its numerical value is very close to $-2M_{K}^{\rm T}$. Therefore the Komar mass associated with the anti-trapping horizon is given by $M_{\rm K}^{\rm AT} \approx M_{\rm K}^{T} - 2M_{K}^{\rm T} = -M_{\rm K}^{\rm T}$, and the minus sign is just right because while the translational Killing field is {\rm future} directed on the trapping horizon \textbf{T}, it is \emph{past} directed   on the anti-trapping horizon \textbf{AT}! (See the (blue) arrows in Fig. \ref{fig:penrose}.) This resolution is another example of the conceptually subtle balance achieved with the choice of quantum parameters (\ref{db-dc}).

To summarize, the Schwarzschild singularity is naturally resolved in the effective theory discussed in Section \ref{s2.1} and region II of Fig. \ref{fig:kruskal} bounded by the singularity to the future is extended to the  singularity free diamond-shaped region shown in Fig. \ref{fig:diamond}, bounded in the past by the trapping horizon and to the future by the anti-trapping horizon. The singularity is replaced by a space-like surface $\T$ that marks the transition between trapped and anti-trapped regions. Curvature scalars achieve their maximum values on $\T$  which are universal to the leading order. Although quantum corrections encoded in the area gap $\Delta$ dominate near $\T$, they decrease rapidly as one moves away and are completely negligible near horizons for macroscopic black holes.  In particular, the radii of the trapping and anti-trapping  horizons are indistinguishable for macroscopic black holes.

\subsection{Summary of LQG Investigations}
\label{s2.3}

As we already noted, the LQG investigations of the Schwarzschild singularity follow the same general steps but differ in the selection of the quantum parameters $\delta_b,\, \delta_c$. Since the Schwarzschild interior is isometric to the Kantowski-Sachs cosmological model, discussions have often focused on issues motivated by cosmological considerations such as the behavior of `scalar factors' and shears, rather than on considerations that are more directly relevant to black holes, in particular properties of the effective geometry that lead to trapping and anti-trapping. We focused on an approach that does \cite{aoslett,aos}. We will now summarize various strategies that have been used to fix $\delta_b,\, \delta_c$ and results they led to. Since our goal is only to present a cohesive picture of the overall status through comparison of results, the discussion will be rather brief; details can be found in the original papers listed in the bibliography.

By and large, these strategies fall into three categories:\\
(i) The parameters are chosen to be constants. These approaches are often referred to as the $\mu_o$-type schemes because they mimic the strategy of using constant values for the quantum parameter $\mu$ used in LQC \cite{abl}. Here, the curvature operator is defined using holonomies of the gravitational connection around plaquettes and shrinking them till the \emph{coordinate} area they enclose equals the area gap $\Delta$;\\
(ii) The parameters are chosen to be phase space functions, using physical considerations. These approaches are often referred to as the $\bar\mu$-type schemes, named after the strategy of selecting the quantum parameter $\mu$ in LQC \cite{aps} in which the curvature operator is defined by shrinking the plaquettes till the \emph{physical} area they enclose equals $\Delta$;\, and,\\
(iii) The parameters are chosen to be phase space functions that are constants of motion on the effective dynamical trajectories. The strategy used in the last two sub-sections falls in this class.

The earliest investigations \cite{ab,lm,cgp} used strategy (i); technically it is the simplest to implement. Here the quantum parameters were set to $\delta_b= \delta_c = 2\sqrt{3}$ using `square' plaquettes  in coordinates adapted the symmetries. Predictions of the resulting effective theory were analyzed in detail in \cite{cs}. The singularity is again resolved and replaced by a 3-surface at which the symmetry 2-spheres attain the minimum area. However, physical quantities such as the minimum value of the radius and the radius of the anti-trapping horizon now depend on the infrared cutoff $L_\circ$. Another limitation is that quantum effects can become significant even in the low curvature region near the horizons.

In the approaches \cite{bv,dc} based on strategy (ii), the quantum parameters were fixed by mimicking the successful $\bar\mu$ strategy from LQC. One again adapts the plaquettes to the symmetries of the problem, but shrinks them till the \emph{physical} area they enclose is $\Delta$. Therefore the plaquettes themselves now depend on the phase space point under considerations and change under time evolution. As a consequence, quantum parameters  are specific phase space functions that are not constant along dynamical trajectories: $\delta_b = \Delta/p_c$ and $L_\circ^2\,\delta_c^2 = (L_\circ^2\, p_c \Delta)/p_b^2$. In these definitions, the dependence of $L_\circ$ is exactly the one that is needed to assure that physical results are independent of the fiducial choice of $L_\circ$. This is a significant improvement over results from strategy (i). However, a technical complication arises because $\delta_b$ depends on $p_c$ and $\delta_c$ on $p_c$: the equations in the $b$ and $c$ sectors no longer decouple. Consequently, it has not been possible to write down analytic solutions and all explorations to date have been performed numerically. These calculations show that the framework has two types of limitations. First, as in (i), there are  large deviations from the classical theory even when the curvature is low. Second, when one evolves beyond the transition surface, the dynamical trajectory enters a region of the phase space where the metric 2-spheres have area that is less than the area gap $\Delta$, making the scheme internally inconsistent. Perhaps not surprisingly, then, some of the properties of the extended space-time are difficult to understand physically.

Strategy (iii) was first adopted to improve on this situation by making $\delta_b,\, \delta_c$ phase space functions that remain constant along dynamical trajectories \cite{cs,oss}. Then the considerations of the first part of Section \ref{s2.1} are applicable, the $b$ and the $c$ sectors separate, dynamical trajectories can be written down analytically, and $m_b$ and $m_c$ are constants of motion. In the first investigation, $\delta_b$ and $\delta_c$ were chosen by dimensional considerations and by taking into account the fact that it is only the combination $(L_\circ  \delta_c)$ that is invariant under the change of the infrared cutoff\, $L_\circ$. The simplest expressions satisfying these requirements were then selected, $(\delta_b)^2 := \Delta/4 m^2$ and $L_\circ (\delta_c)^2 = \Delta$,\,  without the considerations of plaquettes and holonomies of the gravitational connection around them \cite{cs}. The physical results are now invariant under rescalings of $L_\circ$ as desired. There is again a transition surface $\T$ that separates the trapped and anti-trapped regions, and the quantum corrected space-time is a diamond bounded by a trapping horizon in the past and an anti-trapping horizon in the future. Furthermore, unlike the $\mu_o$ and $\bar\mu$-type schemes, quantum corrections are small in regions near the horizons where the curvature is low. However, detailed examination revealed two limitations. First, at the transition surface the Kretchmann scalar of (initially) macroscopic black holes now goes as $1/m$; whence it \emph{decreases} as the mass of $m$ increases. Therefore for astrophysical black holes, large quantum corrections at the heart of the `bounce' at $\T$ occur at low curvature. A second counter-intuitive result is involves `mass inflation' across $\T$. The radius $r_{\rm AT}$ of the horizon in the future of $\T$ now goes as\, $r_{\rm AT} = (r_{{}_{\rm T}})\times (r_{{}_{\rm T}}/ \lp)^3$. Therefore, if the initial black hole has solar mass with $r_{{}_{T}} = 3$km,\, one has $r_{\rm AT} \approx 10^{93}$Gpc! The physical mechanism responsible for this huge magnification has remained unclear. Therefore, subsequently, more general choices of the quantum parameters were explored by introducing new dimensionless constants $\alpha$ and $\beta$, setting $(\delta_b)^2 := (\alpha^2\, \Delta)/4 m^2$ and $L_\circ (\delta_c)^2 = \beta^2 \Delta$ and varying $\alpha$ and $\beta$ to ensure $r_{\rm AT} \approx (r_{{}_{\rm T}})$ for large black holes. Two choices satisfying this condition were found numerically and one analytically. The analytic expression implies that the leading term in Kretchmann scalar at $\T$ is not universal but grows rapidly with $m$ as $m^4/\Delta^4$.

The expressions (\ref{db-dc}) used in Section \ref{s2.2} to discuss results also fall under strategy (iii). However, now the quantum parameters $\delta_b$ and $\delta_c$ are obtained using certain plaquettes, holonomies around which are used to define the curvature. These plaquettes are tailored to the symmetries of the problem, and enclose physical area $\Delta$ as in the strategy (ii). The key difference is that these loops are restricted to lie on the transition surface $\T$ \cite{aoslett,aos}. Since each dynamical trajectory intersects $\T$ once and only one, the prescription is unambiguous and, by construction, makes $\delta_b$ and $\delta_c$ constants of motion. This choice automatically leads to the result $r_{\rm AT} \approx (r_{{}_{\rm T}})$, without recourse to any additional free parameters (such as $\alpha,\, \beta$ discussed above). Furthermore, now the transition surface necessarily lies in the region where curvature is Planckian and the leading terms in the expressions of all curvature invariant are universal.

As this discussion shows, the task of choosing appropriate quantum parameters $\delta_b, \delta_c$ is a very subtle. While is it not difficult to make a `reasonable' choice that resolves the singularity, the resulting quantum corrected geometry has to satisfy several non-trivial constraints to be physically admissible. Over the years, several choices have been proposed but the subsequent careful scrutiny by the LQG community showed that they lead to results that are physically unsatisfactory in one way or the other. The choice discussed in the last two subsections passes all the checks known to date. While this is satisfying, the analysis is still incomplete in one respect. In LQC the effective equations could be derived systematically starting from the operator equations of the quantum theory, showing that there are states that remain sharply peaked even in the deep quantum regime, and using expectation values of observables in these states \cite{jw,vt} (and Section V of \cite{asrev}). Thus the LQC effective equations encode the dynamics of the peaks of these wave functions.  For black holes, the successful LQC techniques have been used in conjunction with an extended phase space framework (introduced in \cite{aos}) to arrive at the desired operator equations and to select physical states in \cite{menaetal}. A systematic derivation of effective equations from this quantum theory remains an interesting open issue in LQG.

\section{The Schwarzschild exterior}
\label{s3}

For the discussion of singularity resolution, it suffices to consider just the region II of Fig. \ref{fig:kruskal}. Therefore, initially the focus on LQG investigations was on this region. However, for a complete understanding of the quantum corrected space-time, one also has to connect the effective space-time geometry of region II to that of region I. In Section \ref{s3.1} we present an approach to carry out this task. Section \ref{s3.2} summarizes the properties of the near-horizon quantum corrected geometry it provides,\, and \ref{s3.3} discusses the asymptotic structure of the effective space-times. As expected, for macroscopic black holes the near horizon geometry exhibits physically expected features because quantum corrections are small there. In the asymptotic region, on the other hand, this effective geometry has an unforeseen feature: while the quantum corrected metric is asymptotically flat in a precise sense, the approach to flatness is weaker than what one might have a priori expected. We will discuss this issue and summarize its current status in Section \ref{s6}.

\subsection{The underlying framework}
\label{s3.1}

Recall that the analysis of the Schwarzschild interior was greatly facilitated by the fact that this region is foliated by \emph{homogeneous, space-like} slices. The exterior region on the other hand does not admit such a foliation. However, the four Killing fields do provide a natural foliation of this region by homogeneous, \emph{time-like} slices. Indeed the textbook derivation of the classical Schwarzschild metric can be interpreted as solving the `evolution equation' in the $r$ direction together with the `Hamiltonian' constraint on the $r ={\rm const}$ homogeneous slices,  mirroring the procedure used in the Schwarzschild interior (or, Kantowski-Sachs space-times). The main difference is that the signature of the intrinsic 3-metric on the homogeneous slices is now -,+,+ rather than +,+,+. Therefore in the connection framework one has to change the internal group that acts on the orthonormal triads from SU(2) to SU(1,1).\footnote{This strategy of using time-like 3-manifolds to specify fields and then `evolving' them in space-like directions was proposed and pursued in \cite{hlkn} for the Hamiltonian framework of full LQG. As discussed there, in the full theory one encounters certain non-trivial technical difficulties associated with the fact that SU(1,1) is non-compact. These issues do not arise in the homogeneous context discussed here.}
The generators $\tau_i$ that provide a basis for the Lie algebra of SU(2) are now replaced by $\t\tau_i$ that constitute a basis for the Lie algebra of SU(1,1). The relation between the two is
\be \label{conversion}
\t\tau_{1} = i \tau_{1}, \quad \t\tau_{2} = i \tau_{2}, \quad{\rm and} \quad \t\tau_{3} =  \tau_{3}. \ee
Hence, for exterior region we can choose our basic variables to be
\ba \label{AEt}
A^i_a \, \t\tau_i \, \d x^a \, &=& \, \f{\t{c}}{L_{o}}  \, \tau_3 \, \d x +  i\t{b} \, \tau_2 \d \theta
- i\t{b}\, \tau_1 \sin \theta \, \d \phi + \tau_3 \cos \theta \, \d \phi,\nonumber\\
E^a_i \, \t\tau^i \partial_a \, &=&  \, \t{p}_c \, \tau_3 \, \sin \theta
\, \partial_x + \f{i\t{p}_b}{ L_o} \, \tau_2 \, \sin \theta \,
\partial_\theta -\f {i\t{p}_b}{ L_o} \, \tau_1 \,  \partial_\phi .
\ea

Comparison with (\ref{AE}) reveals that one can arrive at solutions to the  `constraint' and `evolution' equations in the exterior region simply by using the substitutions\,\,\\
\centerline{ $b \to i\t{b}, \,\,\,  \, p_{b} \to i \t{p}_{b} \qquad {\rm and} \qquad  c\to \t{c},\,\,\, p_{c} \to \t{p}_{c}$}
in the solutions of the interior region. Indeed, one can explicitly check that if one makes these substitutions in the classical solutions (\ref{conf}) and (\ref{momenta}), one obtains the Schwarzschild metric in the exterior region. Therefore we can use these substitutions in the solutions (\ref{eq:c}), \, (\ref{eq:b})\, (\ref{eq:pb})\, to the effective equations in the interior to obtain the desired dynamical trajectories in the exterior region, $T >0$. They yield
\ba \label{eq:ct}
\tan \Big(\f{\delta_{c}\, \t{c}(T)}{2} \Big)&=&  \f{\gamma L_o \delta_{c}}{8 m} e^{-2 T},\qquad
\t{p}_c(T) = 4 m^2 \Big(e^{2 T} + \f{\gamma^2 L_o^2 \delta_{c}^2}{64 m^2} e^{-2 T}\Big) ,\\
\label{eq:bt}
\cosh \big(\delta_{b}\,\t{b}(T)\big) &=& \t{b}_o \tanh\left(\f{1}{2}\Big(\t{b}_o T + 2 \tanh^{-1}\big(\frac{1}{\t{b}_o}\big)\Big)\right),
\ea
where $\t{b}_o = (1 + \gamma^2 \delta_b^2)^{1/2}$\,\, $\delta_b,\,\delta_c$ are given by (\ref{db-dc}) as in Section \ref{s2},\, and,
\be\label{eq:pbt}
\t{p}_b(T) = - 2 \, \f{\sin (\delta_{c}\, \t{c}(T))}{\delta_{c}} \f{\sinh (\delta_{b}\, \t{b}(T))}{\delta_{b}} \f{|\t{p}_c(T)|}{\gamma^{2 }-\f{\sinh^2(\delta_{b}\, \t{b}(T))}{\delta_{b}^2}}\, .
\ee
Thus, the explicit solutions in the $c$-sector have the same form as their counterparts (\ref{eq:c}) in the interior region ($T <0$) while in the $b$-sector the trigonometric functions of $(b\delta_b)$ are replaced by their hyperbolic analogs. Details of derivations and a discussion of the comparison between the classical and effective descriptions of the exterior region can be found in \cite{aos}.

Let us conclude by specifying space-time geometry in the exterior region. The translational Killing field --which is time-like in the exterior region-- is still given by $\partial/\partial x$ and $T$ is a radial coordinate that vanishes on the horizon and is positive in the exterior region. For $T>0$, the effective metric is given by
\be \label{tg}
\t{g}_{ab} \d x^{a} \d x^{b} =  - \f{\t{p}_b^2}{|\t{p}_c| L_o^2} \d x^2 + \f{\gamma^{2} |\t{p}_{c}|\, \delta_{b}^{2}}{\sinh^{2} (\delta_{b}\t{b})} \d T^2+ |\t{p}_c| (\d \theta^2 + \sin^2\theta \d \phi^2).
\ee
The metric is well-defined in this region and has signature -,+,+,+. It fails to be well-defined at $T=0$ because $b$ and $p_b$ vanish there. However, as we show below, this is just a reflection of the breakdown of the coordinate system.  In the limit $\lp \to 0$ (or, $\Delta \to 0$, keeping $\gamma$ positive), the quantum parameters $\delta_b$ and $\delta_c$ vanish and the metric (\ref{tg}) reduces to the Schwarzschild metric in the exterior region.  Properties of the geometry induced by this effective metric are discussed in the next two subsections.

\subsection{Quantum corrected, near horizon geometry}
\label{s3.2}

In this subsection we will briefly discuss two features of the near horizon geometry: Matching of the effective metric across the horizon and corrections to the Hawking temperature, computed using Euclidean (or rather, Riemannian) geometry. Further details can be found in \cite{ao}.
\medskip

\emph{Matching across horizon $T=0$.} Recall that in the classical theory, although the metric appears to be ill-defined across the horizon, one can introduce Eddington-Finkelstein type coordinates to make its regularity explicit. The same strategy can be adopted at the horizon $T=0$ of the effective metric. As in the classical case, one can ignore the angular part of the metric. Then the relevant 2-metrics in interior and the exterior can be respectively written in the form
\be \rmd S_{2}^{2} = f_{1}(T) \rmd x^{2} - f_{2}(T) \rmd T^{2}; \quad {\rm and} \quad
       \rmd \t{S}_{2}^{2} = -\t{f}_{1}(T) \rmd x^{2} + \t{f}_{2}(T) \rmd T^{2}, \ee
where
\be \label{f1f2} f_{1}(T) = \f{{p}_b^2}{{p}_c L_o^2}, \,\,\,{f}_{2}(T) = \f{\gamma^{2} {p}_{c}\, \delta_{{b}}^{2}}{\sin^{2} (\delta_{{b}}{b})}\, \quad {\rm and} \quad
\t{f}_{1}(T) = \f{\t{p}_b^2}{\t{p}_c\, L_o^2}, \,\,\,\t{f}_{2}(T) = \f{\gamma^{2} \t{p}_{c}\, \delta_{\t{b}}^{2}}{\sinh^{2} (\delta_{\t{b}}\t{b})}\,.
\ee
 %
%,
\nopagebreak[3]\begin{figure} [h] %\bfig
\begin{center}
\includegraphics[width=0.8\textwidth]{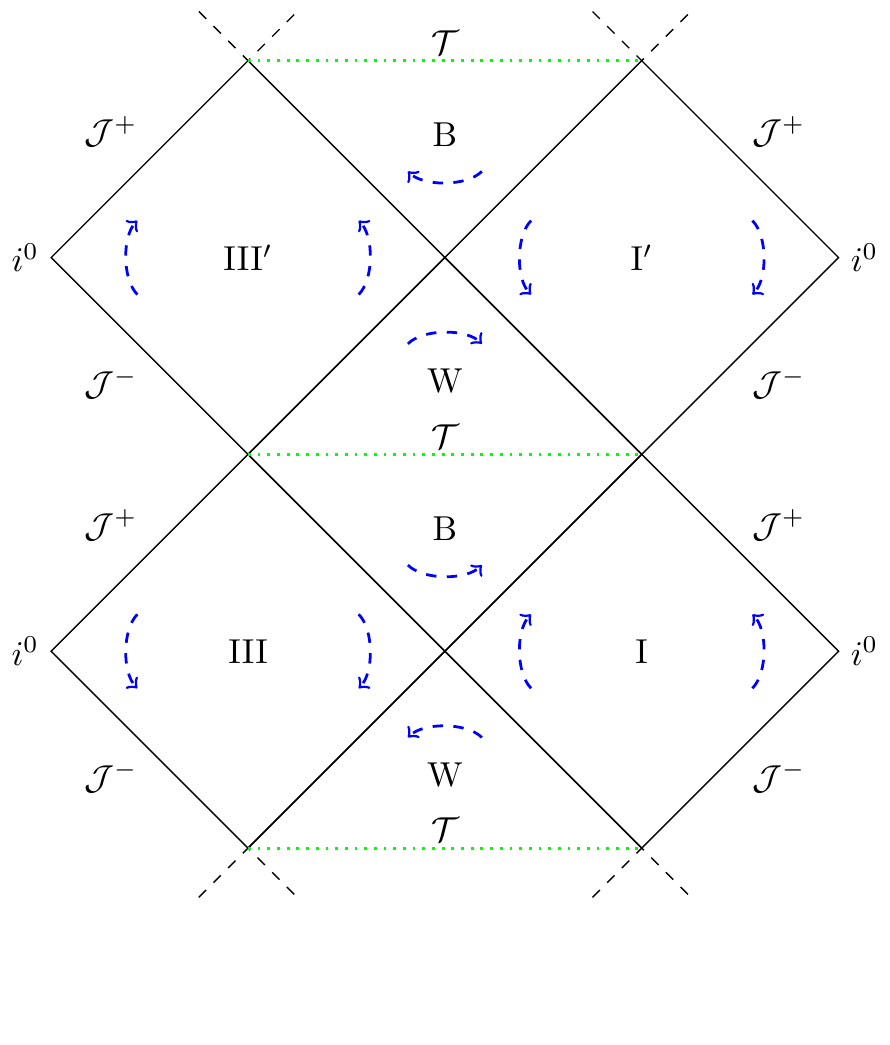}
\end{center}
\caption{\footnotesize{The Penrose digram of the quantum extended Kruskal space-time in the $x, T$ plane. Arrows show the orientation of the static Killing fields. Since the effective metric is at least $C^2$ across Killing horizons, space-time continues indefinitely into the future and the past. Successive Killing horizons are trapping and anti-trapping. Each diamond shaped region they bound is divided by a space-like transition surface $\T$ that separates a trapping region (that lies to the past of $\mathcal{T}$) and an anti-trapping region (that lies to its future). Thus, the region $B$ immediately to the past of $\T$ resembles a black hole interior, and the region $W$ immediately to its future resembles a white hole interior. The area-radii of successive horizons are very nearly equal for macroscopic black holes. Only a part of this extension is relevant to black holes formed dynamically through collapse.}}
\label{fig:penrose}
\end{figure}

As in the Eddington-Finkelstein extension in the classical case, one can approach the horizon from the exterior region. Remembering that the coordinates $(T, x)$ used for the effective metric are the analogs, respectively, of the Schwarzschild coordinates $(r, t)$, one defines an advanced null coordinate $v = x + \t{T}_{\star}$ where
\be \rmd \t{T}_{\star} =  \Big({\t{f}_{2}}/{\t{f}_{1}}\Big)^{\f{1}{2}} \rmd T. \ee
Then the metric in the exterior region becomes
\be  \rmd S_{2}^{2} = \t{f}_{1}\, \rmd v^{2} - 2\, (\t{f}_{1}\, \t{f}_{2})^{\f{1}{2}} \,\,\rmd v\, \rmd T.  \ee
Since $\t{f}_{1}$  vanishes at $T=0$, the space-time metric is well-defined at the horizon with signature -,+,+,+ if and only if $\t{f}_{1}$ is smooth, and $\t{f}_{1} \t{f}_{2}$ is smooth and positive in a neighborhood of $T=0$. This is indeed the case. In particular, \, $\lim_{T\to 0}  \t{f}_{1} \t{f}_{2} = 4m^{2}$. In the standard Schwarzschild coordinates $(r,\,t)$ used in the classical theory, the product is $1$, and since $r= 2m\,e^{T_{\rm class}}$, it is again $4m^2$ in the $(T_{\rm class},\, x)$ coordinates. In this sense the product is the `same' for the classical and the effective metric. The first derivative of $\t{f}_1$ differs from its classical values by terms of the order ${0}(\epsilon_m)$ where $\epsilon_m = {(\gamma^2 L_0^2 \delta_c^2)}/{64 m^2}$ and the second derivative by terms of the order ${0}(\epsilon_m,\, \delta_{b}^2)$. For the metric coefficient $(\t{f}_{1}\, \t{f}_{2})^{\f{1}{2}}$, they are given by $2m$ and $m\big(2+\gamma^{2}\delta_{\t{b}}^{2}\big)$, respectively. If one approaches the horizon from the interior, one finds that the limits of $f_1$ and $(f_1 f_2)^{\f{1}{2}}$ and their first two derivatives exist and match with those coming from the exterior.  Thus, the effective metric is (at least) $C^{2}$ across the horizon $T=0$. Furthermore, the corrections to the metric coefficients are negligible for macroscopic black holes.

In summary, although the effective 4-metric is constructed in the interior region $T<0$ using spatial homogeneity of a space-like foliation and in the exterior region $T>0$ using temporal homogeneity of a time-like foliation, and the $x$ coordinates becomes ill-defined at the horizon, as in the classical theory, there is a well-defined Eddington-Finkelstein type chart $(v, T)$ in which {$\rmd S^2_2$} is well-defined also at $T=0$. Therefore the effective metric can be extended across both the future and past horizons as in the classical Kruskal case shown in Fig.~\ref{fig:kruskal}. Furthermore, since the singularity is resolved, one can extend the metric also across the new, anti-trapping horizons shown in Fig.~\ref{fig:diamond}. One can continue these extensions to arrive at the Penrose diagram of \ref{fig:penrose} which extends indefinitely to the future and to the past. \bigskip

\emph{Quantum corrections to the Hawking temperature.} In the classical theory one can arrive at the Hawking temperature by passing to the Riemannian section via wick rotation of the metric in the exterior region. In these considerations, it suffices to restrict oneself to the $r,\,t$ plane where the Riemannian metric $\t{\g}_{ab}$ has the form
\be \label{eucl-metric} \t{\g}_{ab} \rmd x^{a} \rmd x^{b} = \t{f}_{1}(r) \, \rmd t_{\rm E}^{2} + \t{f}_{2} (r) \rmd r^{2}\, \ee
Since the norm of the translation Killing vector $\t{f}_1(r)$ vanishes at the horizon in the Lorentzian section and since the only vector that has vanishing norm is the zero vector in Riemannian signature, the horizon shrinks to a point where the Killing vector vanishes. In a neighborhood of this point, the static Killing field resembles a rotation, whence $t_{\rm E}$ becomes periodic with period $\P$. This `rotational' character of $t_{\rm E}^{a}$ becomes manifest if we set $R = (\t{f}_{1}(r))^{{1}/{2}}$ \, so that the metric on the $r-t_{\rm E}$ plane becomes
\be {\t{\g}_{ab}} \rmd x^{a} \rmd x^{b} = R^{2}\, \rmd t_{\rm E}^{2} + 4 \,\f{\t{f}_{1}\t{f}_{2}}{{(\t{f}_{1}^{\prime})^2}} \, \rmd R^{2}\, . \ee
The requirement that the metric be free of a conical singularity at the point $R=0$ (where the Killing Field vanishes) constrains the period $\P$ of $t_E$ to be
\be \label{period} \P \,=\, \lim_{R\to 0} \, \f{4\pi (\t{f}_{1}\t{f}_{2})^{\f{1}{2}}}{\t{f}_{1}^{\prime}}  \,= \, \lim_{R\to 0}\, \f{4\pi (\t{f}_{1})^{\f{1}{2}} }{||D \t{f}_{1}|| }\,   \ee
where the last step brings out the invariant nature of $\P$ since it involves only the norm $\t{f}_1$ of the Killing field and the norm of its covariant derivative. This periodicity implies that Green's functions satisfying standard boundary conditions in the Riemannian sector have the same periodicity, which is used to endow the temperature $T_{\rm H} = \hslash/(K\P)$ to the black hole through the relation between Lorentzian field theories and their Wick rotated versions \cite{ggmp,fulling}. For the classical Schwarzschild solution, we have $\P = 8\pi m$, which yields
$T_{\rm H} = \hslash/(8\pi\, K\, m)$

This strategy can be directly applied to the effective metric (\ref{tg}) in the exterior region. The Wick rotated, positive-definite metric in the $(r,\, t)$ plane  --i.e., now  in the $(T,\, x)$ plane-- becomes:
\be \label{2metric}
 {\t{\g}_{ab}} \rmd x^{a} \rmd x^{b}\, =\,   \t{f}_{1} (T) \rmd x^{2} + \t{f}_{2} (T) \rmd T^{2}  \quad {\rm with}
 \quad \t{f}_{1} = \f{\t{p}_b^{\,2}}{\t{p}_c\, L_o^2} \quad {\rm and}\quad  \t{f}_{2} =  \f{\gamma^{2}\, \t{p}_{c}\, \delta_{\t{b}}^{2}}{\sinh^{2} (\delta_{\t{b}}\t{b})}\, \nonumber .\ee
The horizon is at $T=0$, where $\t{p}_{b}$ and $\t{b}$ vanish in the effective solution. Regularity of the metric follows from the properties of $\t{f_1}$ and $\t{f}_1 \t{f}_2$ discussed above.  The period $\P$ of (\ref{period}) is now given by $\P = 8\pi m (1+ \epsilon_m)$ where, as before, $\epsilon_m = {(\gamma^2 L_0^2 \delta_c^2)}/{64 m^2}$.
Therefore, the Hawking temperature of the quantum corrected black hole horizon  is
\be \label{temp} T_{\rm H}\,  =\, \f{\hslash}{8\pi K m} \, \f{1}{(1+\epsilon_{m})}\ee
The mass dependent correction $1/(1+\epsilon_{m})$ due to quantum geometry effects is very small for macroscopic black holes. For a solar mass black hole it is of the order of {$\sim 4 \times 10^{-106}$}. Indeed, even for a black hole of $\sim 10^{6} M_{\rm Pl}$, the correction is of the order {$10^{-21}$}. (Because there are inherent approximations in arriving at the effective theory, further extrapolation to even smaller black holes would not be appropriate.)

As discussed in Section \ref{s2}, the quantum corrections  to various curvature invariants are very small near the horizon of macroscopic black holes. The correction $\epsilon_{m}$ to the Hawking temperature provides another facet of that general phenomenon.

\subsection{Asymptotic properties of the effective geometry}
\label{s3.3}

As we saw, the quantum gravity corrections are very small near horizons of macroscopic black holes. Exact  calculations have been done using MATHEMATICA in a (large) neighborhood of the horizon as one recedes outwards and they show that quantum corrections to the geometry become even smaller, as one would expect. However, as one recedes further to asymptotic regions $r \gg 2 m$, the trend does not continue. The main issue is tied with certain subtleties related to asymptotic flatness and the associated Arnowitt, Deser, Misner (ADM) energy that are not widely appreciated and can lead to confusion (for details, see \cite{ao}).

Let us therefore begin by recalling the elementary notion of asymptotic flatness. A given metric $g_{ab}$ is said to be asymptotically flat at spatial infinity  if there exists a flat metric $\mathring\eta_{ab}$ such that in a Cartesian chart  defined by $\mathring\eta_{ab}$, components of $g_{ab}$ approach the components  of $\mathring\eta_{ab}$ at least as fast as $1/r$ as $r \to \infty$, keeping $t, \theta,\varphi$ constant  (where $(t, r,\theta,\varphi)$ refer to $\mathring\eta_{ab}$ ). However, $\mathring\eta_{ab}$ may not be the `obvious' flat metric suggested by the coordinates in which $g_{ab}$ is presented. An obvious example is the 2-dimensional metric $\bar{g}_{ab}$ with the line element $\rmd \bar{s}^{2} = - r^{2} \rmd t^{2} + \rmd r^{2}$. The fact that $\partial/\partial t$ is the Killing vector of the metric suggests that the coordinates $t,\, r$ are `natural', whence one may be led to consider the flat metric $\bar{\eta}_{ab}$ with the line element $\bar{\eta}_{ab} \rmd x^{a} \rmd x^{b} = -\rmd t^{2} + \rmd r^{2}$. One would then conclude that the given metric $\bar{g}_{ab}$ is \emph{not} asymptotically flat because it does not approach $\bar{\eta}_{ab}$. Indeed, this conclusion may be further re-enforced by the fact that the norm of the static Killing field diverges as $r\to \infty$. But not only is  $\bar{g}_{ab}$ asymptotically flat, it is in fact flat because $\bar{g}_{ab}$  is just the Minkowski metric in the Rindler wedge. This example brings out the fact that even a flat metric is generically not asymptotically flat w.r.t. other flat metrics even in the elementary sense! Note, however, that for a given metric $g_{ab}$ to be asymptotically flat, it suffices to find \emph{one} flat metric, say $\mathring{\eta}_{ab}$, to which it approaches; it need not approach a pre-selected flat metric, like $\bar\eta_{ab}$ in the above example. A more subtle example is provided by the Levi-Civita solution to Einstein's equation (known as the `c-metric') \cite{cmetric} that, it turned out, represents the gravitational field of two accelerating black holes \cite{kw}. In this solution, the norm of the Killing field $\partial/\partial t$ also diverges at spatial infinity, and it too seems not to be asymptotically flat in the coordinates it is normally presented in. (This feature led to considerable confusion on whether this space-time admits gravitational radiation.) But the c-metric is in fact asymptotically flat in the standard sense \cite{aatd} (and does admit radiation); the form of the flat metric $\mathring{\eta}_{ab}$ it approaches at infinity is not obvious in the coordinates the c-metric is presented in.

With these preliminaries out of the way, let us return to the effective metric $\t{g}_{ab}$ of Eq. (\ref{tg}) in the asymptotic region and ask if it asymptotically flat, keeping in mind the subtleties discussed above. Now, $b,c, p_b, p_c$ that enter the expression of $\t{g}_{ab}$ are complicated functions of $T$. To make the asymptotic structure transparent, let us first set
\be r_S := 2m, \qquad   r := r_S\, e^T, \qquad {\rm and}\qquad  \epsilon := 1 -b_0 \equiv 1- (1+ \gamma^2 \delta^2_b)^{\f{1}{2}} \ee
and replace $x$ by $t$ so the translational Killing field  is now $\partial/\partial t$ (rather than $\partial/\partial x$).  For macroscopic black holes the dimensionless parameter $\epsilon$ is very small; for example $\epsilon= 10^{-26}$ for a star mass black hole. Let us therefore assume that $\epsilon \ll 1$. Then in the asymptotic region, where ${r_S}/{r} \ll 1$ and $(\gamma^2\,r_S\,\Delta)^{{1}/{3}}/{2r} \, \ll\, 1\,$, the exact expression (\ref{tg}) of the quantum corrected metric simplifies significantly:\, $\t{g}_{ab}\, \approx\, \t{g}^{\circ}_{ab} \rmd x^{a} \rmd x^{b}\, = \, \t{g}^{\circ}_{tt}\rmd t^{2} + \t{g}^{\circ}_{rr} \rmd r^{2} + r^{2}\,  \rmd \omega^{2} \,$, where,
\be \label{tg-asym} \t{g}^{\circ}_{tt} = -\Big(\frac{r}{r_S}\Big)^{2 \epsilon} \,\Big(1-\left(\frac{r_S}{r}\right)^{1+\epsilon} \Big)\qquad {\rm and} \qquad
\t{g}^{\circ}_{rr} = \Big(1-\left(\frac{r_S}{r}\right)^{1+\epsilon}\Big)^{-1}\, .
\ee

Now, since $\t{g}^{\circ}_{tt}$  --and hence $\t{g}_{tt}$--\, diverges as $r\to \infty$, it is clear that the `obvious' metric does not approach the flat metric $\t{\eta}_{ab} \rmd x^a \rmd x^b = - \rmd t^2 + \rmd r^2 + r^2 \rmd \omega^2$.
Therefore, one may be tempted to conclude that $\t{g}^{\circ}_{ab}$  --and hence $\t{g}^{\circ}_{ab}$-- is not asymptotically flat \cite{brahmacomment}. However, as the examples of the Rindler and the c-metric show, the conclusion does not follow. Rather, the question is whether there \emph{exists} a flat metric $\t{\eta}^{\circ}_{ab}$ to which $\t{g}_{ab}$ approaches as\, $r \to \infty$;\, this $\t{\eta}^{\circ}_{ab}$ need not be the `obvious' flat metric $\t{\eta}_{ab}$. \, The answer turns out to be in the affirmative \cite{ao}. To display its form, one has to replace $t$ with $\tau = t (r/r_S)^\epsilon$ (note that $\tau$ agrees with $t$ for $\epsilon=0$). Then, setting $\t{\eta}^{\circ}_{ab} \rmd x^a \rmd x^b = -\rmd \tau^2 + \rmd r^2 + r^2 \rmd \omega^2$ one finds that components of $\t{g}_{ab}$ approach those of $\t{\eta}^{\circ}_{ab}$ as $1/r$, ensuring asymptotic flatness of $\t{g}_{ab}$. As one would expect from this property, all curvature invariants of $g_{ab}$ vanish as $r\, \to \, \infty$.  Furthermore, this fall-off is sufficient to ensure that the ADM energy is well-defined. It can be computed using the spatial Ricci tensor $\t{\mathcal{R}}_{ab}$ using an expression \cite{aaam2} that is often used in the recent geometric analysis literature on the subject (see, e.g.,\cite{schoen}). One finds
\be \label{ERicci}
E_{\rm Ricci} :=  \lim_{{r} \to\infty}\frac{1}{8 \pi G} \oint_{r} \rmd^{2}V\, \,r\, N \,\, \t{\mathcal R}_{ab} \hat r^a \hat r^b \, \equiv M (1 + \epsilon) ,
\ee
where $\rmd^2 V$ is the area element of the  $r = {\rm const}$  2-sphere of integration,  $\hat{r}^{a}$ a unit radial vector, and $M$ is the Schwarzschild mass of the classical solution. Thus there \emph{is} a quantum correction to the Schwarzschild mass, but it is minuscule for macroscopic black holes.

However, the fact that $\t{g}_{ab}$ does not approach the `obvious' flat metic $\t{\eta}_{ab}$ reflects a limitation of its asymptotic behavior: the approach to flatness is not as strong as assumed in the standard treatments of  asymptotics\, (see, e.g. \cite{aaam2})\, because, while the metric components approach their flat space values as $1/r$, not all components of the connection $\t{\nabla}$ defined by $\t{g}_{ab}$ fall-off as $1/r^2$. As a consequence several components of the space-time curvature have weaker fall-offs than in the standard context. In particular, the curvature invariants fall off only as $1/r^4$ rather than $1/r^6$. These deviations from standard asymptotic behavior have some subtle consequences.

Let us illustrate these subtleties with examples. As we just saw, the expression $E_{\rm Ricci}$ of the ADM energy continues to be well-defined, and yields $E_{\rm Ricci} = M( 1+ \epsilon)$. One can also carry out the calculation using the more familiar expression involving the 3-metric, paying attention to the lapse defined by the Killing field \cite{tt}. One then finds $E_{\rm 3-metric} = M$, without any corrections. Similarly one can also evaluate the mass at the horizon using its area $A_{\rm hor}$,\,\, $M_{\rm hor} = (A_{\rm hor}/16\pi)^{1/2}$\, to find $M_{\rm hor} = M (1 + \epsilon_m)$ where $\epsilon_m$ is the mass dependent term that enters the expression (\ref{temp}) of the corrected Hawking temperature we found in Section \ref{s3.2}. For a solar mass black hole $\epsilon_m \approx 10^{-106}$, much smaller than the correction $\epsilon \approx 10^{-26}$ that enters (\ref{ERicci}). All these quantities agree for the classical Schwarzschild solution because the asymptotic fall-off is the standard one \cite{aaam2}. Now, it often happens that notions that agree in a limiting theory (e.g., Newtonian gravity) become ambiguous in a more complete theory (e.g., GR) and are thus replaced by several different notions. It remains to be seen whether these findings associated with the notion of energy are conceptually similar for the transition from GR to quantum gravity, or if they are blemishes that point to a genuine limitation of the effective metric $\t{g}_{ab}$ in the exterior region, that will be cured by a better candidate. As we will discuss in Section \ref{s6}, this issue is under active investigation in LQG.

\section{Quantum geometric effects in gravitational collapse: illustrations}
\label{s4}

In Section \ref{s2}, we saw that the isometry between the Kantowski-Sachs space-time and the Schwarzschild interior allows one to apply tools from LQC to the Schwarzschild spacetime and permits one to study detailed physical implications. However, these studies have an inherent limitation: they can not capture the dynamics of a gravitational collapse, resulting in a black hole. Models of gravitational collapse are significantly richer: in contrast to eternal black holes, one now has a field theory, in which the time evolution of geometry and matter is coupled and governed by non-linear equations \cite{bojowald-sw,chiou-bh,ben-karim,Kelly:2020vac,Kelly:2020lec,gop:2020,han-imp,gop:2021,Husain:2021ojz,zhang,Husain_2022,Giesel:2022rxi}. In this class of models, several investigations have been carried out to understand the resolution of singularities associated with the dynamical collapse of homogeneous dust in Oppenheimer-Snyder scenarios, in which the interior is modeled by a Friedmann, Lema\^itre, Robertson, Walker (FLRW) cosmology \cite{Giesel:2022rxi,Modesto:2006qh, Hossenfelder:2009fc, Tavakoli:2013rna, Bambi:2013caa, Liu:2014kra, BenAchour:2020bdt,Li:2021snn}. This allows the application of LQC techniques for the study of the fate of the classical singularity and yields similar results on non-viability of certain quantization schemes. In particular, it turns out that the `$\mu_o$ scheme' on which early LQC was based --but subsequently ruled out on cosmological viability criteria \cite{Corichi:2008zb,aps}-- has novel limitations in the black hole sector: it does not permit formation of trapped surfaces unless one chooses rather unnatural features of quantum geometry \cite{Li:2021snn}. This is an illustration of the fact that these models can provide valuable insights, despite the limitations associated with their simplicity.

Another category of investigations considers dynamics of shells where the interior regions is usually a patch of Minkowski spacetime, while the exterior is a Schwarzschild geometry. They  allow for the study of black hole formation, modeling the interior of the star as a simple, empty, flat spacetime. At the quantum level, there is considerable literature on this topic (see for eg. \cite{H_j_ek_2003} for a review). To understand quantum geometry effects in this setting, a reduced phase space quantization of thin shells has been performed \cite{Campiglia:2016fzp,Ziprick:2016ogy,Giesel:2021dug}. One of these works shows that the classical singularity is eliminated, where the shell either emerges through a white hole type geometry or tunnels into a baby universe inside the black hole \cite{Campiglia:2016fzp}. Another work
 proposes an effective semiclassical description motivated by LQC quantization techniques for the study of a Lema\^{i}tre-Tolman-Bondi (LTB) spacetime, focusing on the dynamics of the outermost shell of matter \cite{Giesel:2021dug}. Here, the singularity inside the black hole is resolved. Moreover, after black hole formation, matter bounces, eventually `evaporating' the black hole and dispersing towards infinity. There are also studies that focus their attention to the search of an effective constraint algebra that is free of anomalies, and include the so-called `inverse triad corrections' \cite{Bojowald:2008ja, Bojowald:2009ih}, and `holonomy corrections' \cite{Alonso-Bardaji:2021tvy,Alonso-Bardaji:2021yls}. %with the latter providing an anomaly free description.
 %. For a recent review see \cite{Malafarina:2017csn}.
Finally, there have been studies to understand quantum geometric effects on critical phenomena in the scalar field collapse discovered by Choptuik \cite{Choptuik:1993} in classical GR \cite{Husain:2004yz, Husain:2008tc, Ziprick:2009nd, Kreienbuehl:2010vc,Benitez:2020szx,Benitez:2021,Benitez:2022}.

Given the richness and complexities of the underlying physics, at the present stage these attempts aim at providing insights on specific aspects of the problem, rather than a complete picture.
To illustrate the overall status we will discuss two concrete examples in some detail: the dust collapse scenario, and the critical collapse of a scalar field. The first category of results focus on singularity resolution and therefore use horizon penetrating coordinates. On the other hand, in the second category the focus is primarily on the exterior region, whence it suffices to use coordinates that cover only that part of the space-time. These examples are complementary in the following sense. In the first category, geometry is treated quantum mechanically to start with, and induces quantum effects on matter via field equations. In the second category, to begin with only matter is treated quantum mechanically, and subsequently quantum features descend on geometry from matter, again through field equations.

\subsection{Dust field collapse models}

In this subsection, we consider a few recent investigations \cite{Kelly:2020vac,Kelly:2020lec,Husain:2021ojz,Husain_2022} that illustrate the quantum modifications of classical dynamics. They use a reduced phase space quantization with certain gauge fixing conditions in spherically symmetric space-times, minimally coupled to an inhomogeneous dust field.  The focus is on the family of spherically symmetric Lema\^itre–Tolman-Bondi (LTB) spacetimes, and its sub-family of Oppenheimer-Snyder (OS) models where the dust field is homogeneous. The approach is inspired  by the `improved dynamics' strategy of LQC. In these models the matter sector --dust-- is not quantized but its dynamics is deeply influenced by the quantum nature of underlying geometry, once it enters the high curvature regime.

The metric of LTB space-times is given by \cite{Husain:2021ojz,Husain_2022}
\be\label{eq:ds-ltb}
{\rmd}\,s^2_{\mathrm{cl}} = -N^2 {\rmd}\, t^2+ \frac{E^\varphi(t,x))^2}{E^x} \left({\rmd}\,x +N^x_{\mathrm{cl}}(t,x) \,{\rmd}\,t\right)^2+ x^2 {\rmd}\, \Omega^2 ,
\ee
where $x\in[0,\infty)$ is the radial coordinate and $\varphi$ is the azimuthal coordinate in spatial slices. Let us restrict ourselves to the `marginally bound case' where the spatial slices are flat. In the Hamiltonian framework, one can gauge fix the momentum (or, diffeomorphism) constraint by setting $E^x =x^2$. Preservation of this gauge-fixing condition in time determines the shift $N^x_{\mathrm{cl}}$ in terms of the canonical variables: $N^x_{\mathrm{cl}}(t,x) = - N(K_\varphi(t,x)/\gamma)$\, where $K_\varphi$ is the momentum conjugate to $E^{\varphi}$. (Because the spatial slices are flat, $K_\varphi$ equals the connection component $A_\varphi$.) One can fix the lapse function $N$ without loss of generality; let us set $N=1$ so that $t$ represents proper time. The Hamiltonian constraint relates these geometric variables to the matter density and determines evolution equations for $E^\varphi, K_\varphi$ through Poisson brackets \cite{Husain_2022}.

To pass to the effective theory, one sets $\beta(t,x)=({\sqrt\Delta}/{x})\,K_\varphi(t,x)$ and, motivated by known results in LQC, one makes the ansatz:
\be \label{q-shift}
N^x(t,x) = - \, \f{x}{\gamma \sqrt\Delta} \sin\left( \beta(t,x)\right) ~ \cos\left( \beta(t,x)\right)
\ee
(so that, in  the limit area gap $\Delta \to 0$ (keeping $\gamma >0$), we recover the classical shift $N^x_{\rm cl}$). In the Painlev\'e-Gullstrand like coordinates (for unit lapse),  $E^\varphi$ is time independent, given by $E^\varphi (x,t) = x$. The Hamiltonian constraint and the evolution equation for $K_\varphi$\, --which is now encoded in $\beta$--\, are non-trivial:
\be \label{q-density}
\rho = \f{1}{8 \pi G \gamma^2 \Delta \, x^2} \, \partial_x \left( x^3 \sin^2 \beta \right)\quad {\rm and} \quad
\partial_t \beta = - 4 \pi G \gamma \sqrt\Delta \,\, \rho,  .
\ee
As mentioned earlier, $\rho$ is a classical field throughout this analysis; nonetheless it now acquires an upper bound because of its coupling to quantum geometry.

Within this family of LTB spacetimes, it is interesting to analyze the subfamily of OS solutions, those in which the energy density is homogeneous. The star is bounded by the surface $x=L(t)$, outside of which  $\rho(t)$ vanishes and inside of which $\rho(t)$ is a positive constant for each $t$. Thus, there is a \emph{finite} discontinuity in $\rho$
all along the boundary $x=L(t)$. Eq. (\ref{q-density}) implies that $\beta$ is continuous across the boundary but its time derivative has a finite discontinuity there. One can now solve for the function $\rho(t)$ to obtain
\be \label{L-cont}
\rho(t) = \f{3 G M}{4 \pi \big(L(t)\big)^3} \quad  \hbox{\rm for \, $x < L(t)$}, \qquad {\rm and} \quad  \rho(t) =0 \quad \hbox{\rm for\,  $x > L(t)$}.
\ee
The form of $\rho(t)$ inside the star immediately implies an interesting relation that is reminiscent of the quantum corrected Friedmann equation of LQC \cite{aps}:
\be \label{lqc-fried}
\left( \f{\dot L}{L} \right)^2 = \f{8 \pi G}{3} \rho \left(1 - \f{\rho}{\rho_c}\right),
\ee
for\, $x(t) < L(t)$,\, where $\rho_c = 3/8\pi G \gamma^2\Delta$, is again a \emph{universal constant}.\, (A similar equation of motion for the homogenous dust collapse was obtained in Refs. \cite{Tavakoli:2013rna, Li:2021snn, Giesel:2021dug, Giesel:2022rxi}.)\, At the bounce, one has\, $L_{\rm bounce} = (2GM\gamma^2 \Delta)^{1/3}$;\, the value of the radius at which the bounce occurs grows linearly with the mass of the star. In particular, while the density at the bounce is of Planck scale irrespective of the mass of the star, for macroscopic black holes, the radius at the bounce is not. For a solar mass black hole, for example, $L_{\rm bounce} \approx 10^{13} \lp$. This distinction is a robust feature of LQG.

Since the bounce of the effective theory replaces the classical singularity, one might expect the subsequent dynamics to display richer structure. This is indeed the case. Soon after the bounce, $\beta(x,t)$ develops a discontinuity at the boundary. Therefore, it follows from (\ref{q-density}) that $\rho(x,t)$ acquires a new term that is proportional to the delta distribution $\delta(L(t) -x)$. Consequently, after the bounce the evolution equations have to be solved in the distributional sense; one has weak solutions that solve integral equations obtained by integrating the evolution equation w.r.t. $x$. When the shock wave meets the dynamical horizon \cite{ak2,akrev,boothrev}, it ceases to be a trapping horizon. Taking this instant of the time as the end of the black hole, one can calculate its life time as the proper time interval, \emph{measured  by a distant observer}, between the instant of formation of the dynamical horizon and its disappearance. One finds:
\be\label{eq:Tlife}
T_{\rm lifetime} \sim \frac{8\pi G^2M^2}{3\gamma \sqrt{\Delta}}.
\ee
Although in the above discussion we used the OS solutions to obtain this result, the scaling $T_{\rm lifetime}\, \propto\, M^2$ is more general in LQG. For example, it holds also for shell collapse and the collapse of inhomogeneous dust (up to corrections linear in $M$) \cite{Husain_2022}. This life-time contrasts with the suggestions of  $T_{\rm lifetime}\, \propto\, M$ that have appeared in the literature \cite{Barcelo:2014npa,Barcelo:2014cla,Barcelo:2015uff,Carballo-Rubio:2019nel}, motivated by general quantum gravity considerations but based on less detailed arguments. This possibility is ruled out by the LIGO discoveries of black hole mergers. However, even with the $M^2$ scaling, one is led to the some surprising conclusions. Recall first that the life time of the black hole due to Hawking radiation goes as $M^3$. Therefore, if $T_{\rm lifetime}\, \propto \, M^2$ were to be a firm prediction of a fully developed quantum gravity theory, one would have to conclude that the Hawking evaporation process is physically unimportant since the black hole would disappeared before there is significant Hawking radiation. Secondly, from an astrophysical standpoint, one knows that black holes were formed quite early in the history of the universe. If there were any that  formed with, say, lunar mass, they would have disappeared and left us a signature of the shock wave accompanying the bounce. It is more likely that the $M^2$ scaling will be modified by more complete analyses in the future. For example, the shift is chosen using an educated prescription and not arrived at using some fundamental  principles. In fact, recent investigations indicate that this prescription differs from the one that arises from considerations of dynamical stability of the effective gauge fixing conditions under the effective dynamics generated by the `polymerized' canonical Hamiltonian \cite{Giesel:2021rky}. The usefulness of the current LQG investigations lies precisely in the fact they provide strong and concrete motivation to make the models more and more realistic.

\subsection{Quantum geometric effects in the critical phenomena}

In the classical theory, there are two possible fates for the
 gravitational collapse of a spherically-symmetric, minimally coupled, massless scalar field depending on the initial data. One possible end state is that the field collapses to form a black hole, and the other is that the field disperses to infinity. One can label each family of initial data of the field by suitable parameters $\ulp$, such that for $\underline{p} > \ulp^*$ the collapse leads to a black hole, and for $\underline{p} < \ulp^*$ no black hole forms, i.e., the collapsing scalar  field eventually disperses towards infinity. For $\underline{p}\simeq \ulp^*$, it is possible to form black holes through a second order phase transition with masses as close to zero as desired \cite{Choptuik:1993}.

More precisely, Choptuik demonstrated that the mass of the black hole depends on the difference $(\ulp - \ulp^*)$ via a universal power law  $m_{\mathrm{BH}}\, \propto\,\left|\underline{p}-\underline{p}^*\right|^{{\beta}}$,\, and there exists a discrete self-similar behavior for $\ulp = \ulp^*$. It turns out that $\beta \approx 0.37$ is a universal exponent which is independent of the initial data. Further investigations have brought out a finer structure over and above this power law relation \cite{Hod:1996az}. Due to the  discrete self-similarity one can numerically
observe echoes with a period whose ratio with $\beta$ determines the periodicity in the fine structure. Due to the scale invariance of the underlying equations there is no mass gap for the formation of black holes in the classical theory; black holes can form with arbitrarily small mass.

It is natural to ask: How does this universal phenomenon change when modifications due to quantum geometric effects are included? In LQG investigations of such models, the quantum modifications to the gravitational sector have different origins. The first possibility is to replace the inverse powers of triads using a classical identity to write them as Poisson brackets between holonomies of the gravitational connection and the triads, and then passing to the quantum theory  by replacing the Poisson brackets with commutators \cite{Thiemann:1996aw}. These quantum corrections are often referred to  as `inverse triad modifications'. The second possibility, explained in Section\ref{s2}, is to express the field strength of the connection using  holonomies around closed loops. These modifications are the ones responsible for the bounce of the background effective geometry. In addition one can also treat the matter sector using a polymer quantization \cite{Ashtekar:2002vh}. While a complete treatment to study the critical behavior of the scalar field including all these effects is yet to be performed, explorations have been carried out to understand the modifications of the critical behavior by including only the inverse triad modifications  in Refs.~\cite{Husain:2004yz, Husain:2008tc, Ziprick:2009nd, Kreienbuehl:2010vc}, and by considering LQG quantization of the scalar field in Refs.~\cite{Benitez:2020szx,Benitez:2021,Benitez:2022}. In all these models one assumes the validity of the effective spacetime description resulting in dynamical equations encoding quantum geometry  modifications. Due to inverse triad effects, the behavior of matter-energy modifies the geometry in such a way that there is no divergence and, as a result, the singularity is tamed  \cite{Bojowald:2005qw}. Since inclusion of these modifications inevitably introduces a length scale, the scale-invariance is broken. With these modifications, critical phenomena is recovered albeit with a mass gap, below which a black hole can not form. The value of this gap is determined by the discreteness scale in quantum geometry \cite{Husain:2008tc}. The existence of mass gap on inclusion of inverse triad modifications can also be seen in a more general collapse of the scalar field \cite{Bojowald:2005qw}.

In contrast, if one considers a quantum scalar field \'a la LQG,  one obtains a set of \emph{scale-invariant} effective equations of motion  \cite{Benitez:2020szx,Benitez:2021,Benitez:2022}. Then the mass gap disappears, allowing one to
study of the effects of `polymer quantization' of the scalar field during the formation of black holes of very small masses. Since this treatment closely mirrors the classical theory and, at the same time, captures `polymerization'effects in the matter sector, we discuss it in some detail. The spacetime line element studied in \cite{Benitez:2021} is given by
\be
{\rmd}\, s^2=-N^2(t,x) {\rmd}\, t^2+\frac{\big(E^{\varphi}(t,x)\big)^2}{E^x(t,x)} \rmd\, x^2 + E^x(t,x)\, \rmd\, \Omega^2,
\ee
where one gauge fixes $E^x = x^2$ to parallel the classical treatment by Choptuik.  Its conjugate variable $K_x(t,x)$ is fixed by the diffeomorphism constraint. The shift vector is determined by demanding preservation of the gauge fixing condition in time. One also uses the gauge freedom to set $K_\varphi(t,x)=0$ to maintain the diagonal form of the metric. The dynamical variables are the triad  $E^{\varphi}(t,x)$ and the lapse function $N$.
With the matter content as a scalar field $(\phi(t,x),P_{\phi}(t,x))$, the effective equations of motion are obtained by `polymerizing' the scalar field via $\phi \rightarrow \frac{\sin(k \phi)}{k}$:
\be\label{eq:lapse}
\frac{N^{\prime}}{N}-\frac{\left(E^{\varphi}\right)^{\prime}}{E^{\varphi}}+\frac{2}{x}-\frac{\left(E^{\varphi}\right)^2}{x^3}=0,
\ee
\be\label{eq:ephi}
\frac{\left(E^{\varphi}\right)^{\prime}}{E^{\varphi}}-\frac{3}{2 x}+\frac{\left(E^{\varphi}\right)^2}{2 x^3}-2 \pi x\left(\frac{\left(P_{\phi}\right)^2}{x^4}+\left(\phi^{\prime}\right)^2 \cos ^2(k \phi)\right)=0,
\ee
\be
\dot{\phi}=\frac{4 \pi N}{E^{\varphi} x}P_{\phi} ,
\ee
\be\nonumber
\dot{P}_{\phi}=\frac{4 \pi x^2}{E^{\varphi}}\left[\left(\frac{3 N E^{\varphi}-x N\left(E^{\varphi}\right)^{\prime}+N^{\prime} E^{\varphi} x}{E^{\varphi}}\right) \phi^{\prime} \cos ^2(k \phi)\right.
\ee
\be\left.+x N \phi^{\prime \prime} \cos ^2(k \phi)-x N k\left(\phi^{\prime}\right)^2 \cos (k \phi) \sin (k \phi)\right],
\ee
 The lapse function can be determined from Eq. \eqref{eq:lapse} (which is obtained by imposing preservation in time of the gauge fixing condition $K_\varphi(t,x) =0$.) Finally, the Hamiltonian constraint
Eq. \eqref{eq:ephi} determines the triad $E^\varphi(t,x)$. (Note that for $k\to 0$, (and expressing $E^{\varphi}(t,x) = xa(t,x)$), one obtains the classical equations of motion of \cite{Choptuik:1993}.) One can see that these effective equations remain invariant under the transformation $x\to cx$ and $t \to ct$ for constant $c$. Hence, there will be no mass gap, as in the classical theory. The coordinate system used here cannot penetrate the horizon. Instead, the collapse of the lapse function, namely $N(t,x)\to 0$, is used to signal the formation of a black hole horizon. Numerical simulations with these equations reveal existence of ``wiggles" and  ``echoes" as in the classical description  \cite{Benitez:2020szx,Benitez:2021}. One finds that this effective theory shares the universality of the
scaling of the mass observed in the classical theory, up to small departures for large values of the `polymer parameter '$k$. The period of the discrete self-similarity seems to be independent of the `polymerization parameter' which  indicates that the polymer effective theory has a critical solution with the same periodicity as in the classical theory.\\

 Let us conclude this section with a few remarks. In the investigations of the Kruskal space-time reported in Sections \ref{s2} and \ref{s3}, detailed analysis of quantum corrections to the geometry and their physical implications was made possible, thanks to the presence of a 4-dimensional symmetry group. Dynamical problems discussed in this section have only spherical symmetry and therefore are much more difficult. Thus, various questions remain unexplored. For instance, the quantization scheme (called `K-quantization' \cite{kevinK,Singh:2013ava}), used in
\cite{Kelly:2020vac,Kelly:2020lec,Husain:2021ojz,Husain_2022,Giesel:2022rxi} to arrive at effective equations governing the dust collapse, is only valid for marginally bound cases. Secondly, there are indications \cite{Giesel:2021rky} that  one may have to revisit the assumptions made while `polymerizing' the Hamiltonian constraint, choices made in `polymerization' of lapse and shift, and the issue of consistency of gauge fixing conditions. Further, the choice of shift vector made in  \cite{Kelly:2020vac,Kelly:2020lec,Husain:2021ojz,Husain_2022} and also in \cite{gop:2021} seems to be problematic from the covariance of the effective geometries \cite{gop:2021}.   Finally, there are also studies where another (`non-polymeric') quantization of these classical models has been studied \cite{Schmitz:2019jct,Leveque,Vaz:2011zz,Kiefer:2019csi,Piechocki:2020bfo,Schmitz:2020vdr,Munch:2020czs,BenAchour:2020gon}. A detailed comparison of both quantization schemes could add clarity on the physical viability and mathematical consistency of these two complementary approaches. Similarly, in the investigations of the critical collapse of scalar field, the role of quantum geometry in the gravitational sector is yet to be included \cite{Benitez:2020szx,Benitez:2021}. If one were to introduce `polymerization'of the gravitational connection as in the models for dust collapse, one will very likely introduce a length scale, breaking the scale invariance and a mass gap would appear as in other works incorporating inverse triad modifications  \cite{Bojowald:2005qw,Husain:2008tc}. In explorations of the critical collapse, a more complete picture, including quantum geometric effects in the gravitational sector, is not yet available. This is an important gap as it is these quantum geometry effects that lead to singularity resolution. Despite such limitations, it is encouraging that these models have already provided new perspectives on how quantum effects can manifest themselves in the \emph{dynamical} process of black hole formation and evolution, in the resolution of the classical singularity, and in critical phenomena.

\section{Black hole evaporation}
\label{s5}

Investigations reported in Section \ref{s4} provide interesting insights into the nature of quantum effects in dynamical situations leading to gravitational collapse. However, because of their underlying assumptions, they cannot address the issue of black hole evaporation. In this section we turn to the LQG investigations of the Hawking process and the associated issue of `information loss'.

\nopagebreak[3]\begin{figure}[b] %\bfig
%\sidecaption[t]
\includegraphics[width=0.35\textwidth]{{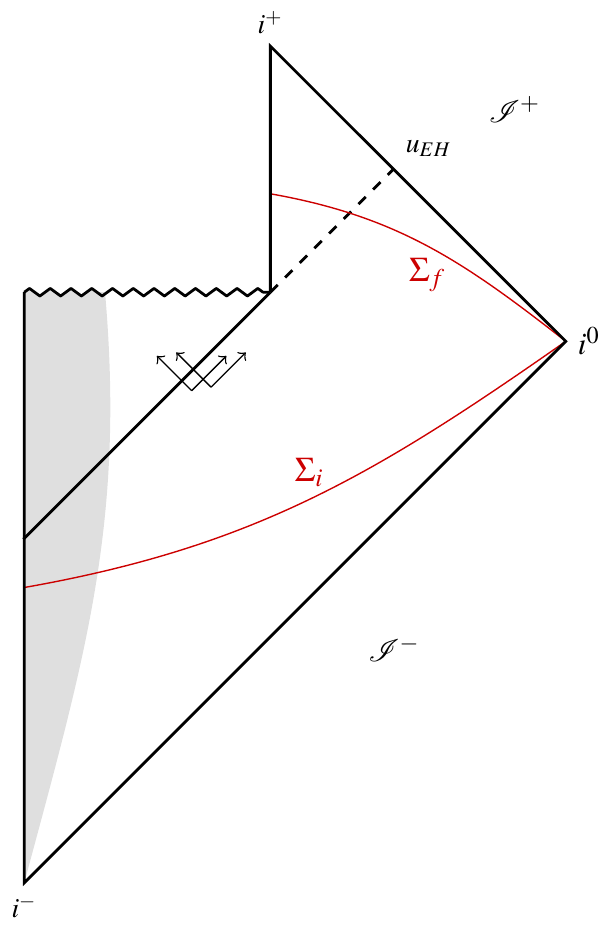}}
\caption{\footnotesize{Commonly used Penrose diagram to depict black hole evaporation, including back reaction. Modes are created in pairs, one escaping to $\scrip$ and its partner falling into the black hole. The dashed line is the continuation of the Event Horizon that meets $\scrip$ at retarded time $u_{EH}$. If this were an accurate depiction, the evolution from $\scrim$ to $\scrip$ would fail to be unitary because the future singularity would act as a `sink of information'.}}
\label{fig:trad}
\end{figure}

In his original discussion \cite{swh} Hawking considered a test, scalar quantum field on a classical space-time depicting gravitational collapse of a spherical star. Heuristic considerations of the inclusion of the back reaction on space-time geometry led to the Penrose diagram of Fig. \ref{fig:trad} that is still widely used. In this diagram $\scri^+$ fails to be the complete future boundary since the singularity is also a part of this boundary. One is then led to the startling conclusion that quantum gravity considerations would force us to generalize quantum physics by abandoning  unitarity \cite{swh2}. However, this line of reasoning has important limitations. The first comes from an elementary observation. For a self-consistent discussion of unitarity, one needs a \emph{closed system}. Thus, the incoming collapsing matter in the distant past has to be represented by \emph{quantum fields}, and the outgoing quantum state in the distant future should refer to the \emph{same} fields. This rather basic point is overlooked in space-time diagram of Fig. \ref{fig:trad} because the asymptotic Hilbert spaces do not include the quantum state of matter in the star. The second issue is more subtle. In much of the discussion on the subject, challenges and paradoxes arise because one assumes that the quantum corrected space-time has an event horizon that encloses a trapped region which is causally disconnected from the asymptotic region. This seems natural from the perspective of the traditional Penrose diagram of Fig. \ref{fig:trad}. However, event horizons are teleological and, as Hajicek pointed out already in 1987 \cite{ph}, they can be shifted arbitrarily, and even completely removed, by changing the space-time geometry in a Planck scale neighborhood of the singularity. Now, there is general consensus that classical GR cannot be trusted in such neighborhoods. Therefore the assumption that the event horizon will persist in quantum gravity has no obvious support. Indeed, LQG considerations suggest that it will not.

In Section \ref{s5.1} we explain how these two issues are addressed in the LQG literature. In Section \ref{s5.2} %and \ref{s5.3}
we summarize the current status of LQG investigations in semi-classical gravity and expectations in full quantum gravity. In broad terms these investigations provide closely related avenues to realize the paradigm introduced in \cite{aamb} based on singularity resolution. \emph{Thus, from LQG perspective, non-singular black holes play a central role in the discussion of the information loss issue.} To anchor the discussion we will use the approach developed in \cite{aa-ilqg,aa-universe}. A complementary discussion can be found in \cite{perez-review}.

\subsection{Setting the stage}
\label{s5.1}

\nopagebreak[3]\begin{figure}[b] %\bfig
%\sidecaption[t]
\includegraphics[width=0.4\textwidth]{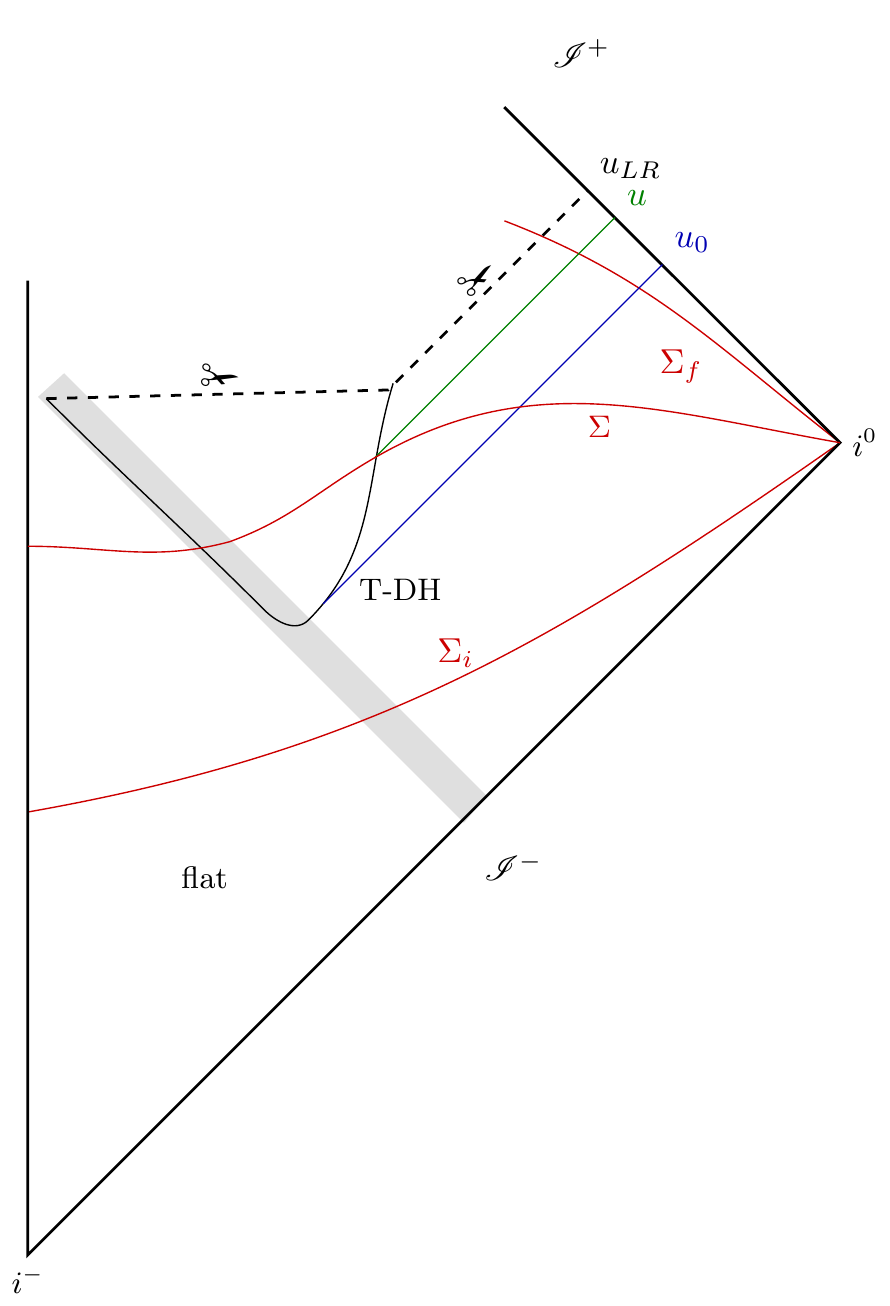}
%
%\caption{\footnotesize
\caption{\footnotesize{Semiclassical space-time: Black hole is formed by gravitational collapse of a pulse of scalar field, depicted by the (gray) shaded region, incident from $\scrim$. A trapping Dynamical horizon \TDH is formed. During the collapse, it is space-like and its area increases (in the outward direction). It becomes time-like during evaporation and its area decreases (in the future direction). Hawking radiation starts in earnest at $u=u_{0}$. The dashed line with scissors that includes the last ray $u =u_{LR}$ represents the future boundary of the semi-classical region.}}
\label{fig:semi}
\end{figure}

A precise formulation of the issue of `information loss' is provided by the question of whether the S-matrix from $\scrim$ to $\scrip$ is unitary which, as we discussed, is relevant only for closed systems. The simplest such system is a massless Klein-Gordon field coupled to gravity. Consider, then, gravitational collapse of a spherically symmetric, massless scalar field $\phi$ from $\scrim$. In the classical theory, if the infalling pulse of $\phi$ is narrow, the collapse is prompt and analysis is not overly contaminated by the details of the pulse profile. The solution has \emph{Minkowski metric} $\eta_{ab}$ to the past of this narrow pulse and a Schwarzschild black hole to its future. It is clear from the lower portion of Fig.~\ref{fig:semi} that the \emph{event horizon first forms and grows in the flat portion of space-time}. The actual collapse could occur billions of years to the future! This is a concrete illustration of the teleological nature of the event horizon ($EH$). In particular, it brings out the fact that the growth of the area of the $EH$ is not tied to any local physical process.

In the quantum theory, the pulse is replaced by a coherent state of the field $\hat\phi$ on $\scrip$. In the semi-classical regime\, --which is expected to be valid in the region in which space-time curvature is much smaller than the Planck scale--\,  one can continue to describe the quantum corrected geometry using a smooth metric. This portion of space-time is depicted in Fig.~\ref{fig:semi}, the region with Planck scale curvature in the future being excised. Let us first focus on this region. The Hawking quanta of the quantum field $\hat\phi$ are emitted in pairs; one escapes to $\scrip$ and its partner falls into the black hole. The quantum state on a Cauchy surface $\Sigma$ of the semi-classical portion of space-time continues to be pure but there is entanglement between the infalling and outgoing quanta. As for geometry, the space-time metric to the past of the infalling pulse continues to be $\eta_{ab}$. But to the future, it is no longer given by the static Schwarzschild solution. The metric is dynamical not only within the pulse but also to its future. Because of its dynamical nature, new structures emerge that are directly relevant to the evaporation process: {\it dynamical horizons}. These are the dynamical analogs of the trapping and anti-trapping horizons of the quantum corrected Kruskal space-time discussed in Section \ref{s2}. They turn out to be more relevant than EHs in discussions of black hole formation and mergers in numerical simulations in classical GR and for the evaporation process in the quantum theory \cite{ak2,akrev,boothrev}.

Let us therefore briefly recall this notion. A {\it dynamical horizon} (\emph{DH}) is a 3-dimensional space-like or time-like submanifold that is foliated by  2-dimensional, surfaces $S$ with 2-sphere topology, such that the expansion of one of the null normals to each leaf $S$ is zero and that of the other null normal is either positive or negative everywhere. Thus, each $S$ is a marginally trapped surface (MTS). In an asymptotically flat space-time,  we can distinguish between the two null normals to $S$. Let us denote by $\l^{a}$ the outgoing null normal and by $n^{a}$ the ingoing null normal. On a black hole type \emph{DH}, the expansion $\Theta_{(\ell)}$ of the outgoing null normal vanishes (it is positive immediately outside and negative immediately inside the MTS), while the expansion $\Theta_{(n)}$ of the ingoing null normal is negative (both outside and inside). Thus, immediately inside a black hole type \emph{DH}, both expansions are negative and we have a trapped region. Therefore, these \DHs are called   \emph{trapping dynamical horizons,} \emph{T-DHs}. A \TDH is space-like when the area of the MTS increases along the projection of $\l^{a}$ on \emph{DH}, i.e. in the outward direction.  In fact, there is an explicit, precise relation between the growth of the area of a \DH and the flux of  energy (carried by matter and/or gravitational waves) flowing \emph{into} it \cite{ak2}. Thus, not only does the second law of black hole mechanics hold on  \emph{T-DHs}  but the growth of the horizon area is \emph{directly} related to local physical processes. This is in striking contrast with the situation for \emph{EHs},  where we only have a qualitative statement of growth in classical GR: Area of $EHs$ cannot decrease. Indeed, it is not possible to directly trace the growth back to the infall of energy locally because, as we just saw, \emph{EHs} can form and \emph{grow} in flat space-time where there is nothing at all falling across it.

During the evaporation process, by contrast, the MTSs on the \TDH shrink, now in response to the local \emph{negative} energy flux across it, and the \TDH is time-like. Recall that in the quantum corrected Kruskal space-time, we also have (white hole type) anti-trapping horizons. But they emerge only when the space-time is extended across the transition surface $\T$ on which curvature is of Plank scale. Therefore one would expect that in dynamical situations, \emph{anti-trapping dynamical horizons} \ATDH would also emerge only when one extends space-time across a transition surface that replaces the classical singularity. This expectation is correct. There is no \ATDH in the semi-classical space-time Fig.~\ref{fig:semi} where the region with Planck scale curvature was excised by hand. However, it is present in the quantum extended space-time depicted in Fig.~\ref{fig:LQG}.

On an \ATDH it is the expansion $\Theta_{(n)}$ of the \emph{ingoing} null normal that vanishes and the expansion $\Theta_{(\ell)}$ of the outgoing null normal is positive. Thus, immediately inside these horizons, both expansions are positive: we have an anti-trapped region. Since it is $\Theta_{(n)}$ that vanishes on any \emph{AT-DH}, it is natural to investigate what happens to the area of the MTSs as one moves along the projection of $n^{a}$ on the \emph{AT-DH}. If the \ATDH is space-like, its area \emph{decreases}  (now in the inward direction) and if it is time-like its area increases (now in the future direction).

Thus, the key differences between \emph{EHs} and \DHs can be summarized as follows. First, \emph{EHs} are teleological and can be located only after one has evolved the metric to infinite future. \DHs by contrast can be located quasi-locally and their properties have direct relation to physical processes at their location. Second, \emph{EHs} are null while \DHs can be space-like or time-like, and become null only when they become `isolated' i.e. there is no flux of energy across them. Third,  nothing can \emph{ever} escape to the `exterior' region from the trapped region enclosed by a black hole type \emph{EH} and nothing can \emph{ever} enter the anti-trapped region bounded by a white hole type \emph{EH}. While there are trapped surfaces immediately inside a \emph{T-DH},  one can send causal signals across a \TDH from inside to outside (see  Fig.~\ref{fig:semi}). Similarly, there are future directed causal curves that traverse an  \ATDH from outside to inside (see  Fig.~\ref{fig:LQG}). Finally, while there is no natural notion of mass and angular momentum for cross-sections of \emph{EHs}, there is one for the canonically defined marginally trapped surfaces on \emph{DHs} which, furthermore lead to the first and second laws of black hole mechanics \cite{ak2}. Discussions of quantum dynamics in LQG focus on \DHs. Much of the confusion about the evaporation process and `purification' of the quantum state melts away once EHs are deemphasized.

\subsection{Black hole evaporation in LQG}
\label{s5.2}

LQG investigations of the semi-classical part of the quantum corrected space-time are based on Fig.~\ref{fig:semi} and, although some of the detailed calculations are still in progress, the overall understanding of structures in this space-time is quite satisfactory at a conceptual level. To understand the structure of the future of this region one needs full quantum gravity and, as in every other approach, several questions remain. But there is a general consensus on a majority of issues. In this subsection we summarize this status.\\

\emph{Semi-classical Regime:} Consider a coherent state $\Psi$ of a quantum scalar field $\hat{\phi}$, peaked around an infalling classical pulse on $\scrim$ and undergoing a prompt collapse. Let us suppose that the ADM energy in the incoming state is of a solar mass, $M_{\odot}$. When the radius of the pulse has become sufficiently small, a trapping dynamical horizon \TDH forms. In classical GR, this \TDH would only have a space-like component that grows from zero radius till it has radius of $3$km and then joins on to the null event horizon of the same radius. Once the Hawking radiation starts and the back reaction is included, the black hole shrinks. Initially the process is very slow because the ingoing negative energy flux is \emph{extremely} small. It takes some $10^{64}$ years for the black hole to shrink to lunar mass $M_{\rm moon}$. However, even at the end of this long, adiabatic process, the black hole is \emph{macroscopic}.  Therefore, from our discussion in Section \ref{s2} one would expect the quantum gravity corrections to be sufficiently small for semi-classical considerations to suffice. Let us focus on this phase of evaporation. In this phase, dynamics should be well-described by equations
\be \label{semi-class} G_{ab}^{\rm sc} = 8\pi G_{\rm N}\,  \langle\, \hat{T}_{ab}\, \rangle_{\rm ren} \quad {\rm and} \quad  \Box\, \hat{\phi} = 0\, , \ee
where $G_{ab}^{\rm sc}$ is the Einstein tensor of the semi-classical metric $g_{ab}^{\rm sc}$ and the expectation value of the renormalized stress-energy tensor is computed using the Heisenberg state $\Psi$. The metric $g_{ab}^{\rm sc}$ does include quantum corrections but they are induced by quantum matter (rather than being dictated by the area gap considerations of quantum geometry).

These corrections to geometry are adiabatic and small. But the infalling negative energy flux introduces a qualitative difference in the horizon structure: Now the expanding, space-like branch \TDH of the dynamical horizon joins on, not to a null event horizon as in the classical case, but to the outer, \emph{time-like} branch whose area \emph{decreases} to the future due to the negative energy flux carried by the Hawking `infalling partner modes'. These two branches of the \TDH serve as the past boundary of a trapped region of the semi-classical space-time $(M_{\rm sc},\, g_{ab}^{\rm sc})$ of Fig.~\ref{fig:semi}. During this long adiabatic process of $\sim 10^{64}$ years, pairs of Hawking quanta are continually created, one going to $\scrip$ and its partner falling into the trapped region.  These modes will be entangled whence, if one uses the usual observable algebra based just at $\scrip$,  the state would seem mixed, close to a thermal state. %(For caveats on departure from thermality, see, e.g.  \cite{anderson}).

The issue of unitarity leads one to ask: When will the correlations be restored? For this to happen, the partner modes would have to emerge from the trapped region and propagate outward, restoring correlations at $\scrip$ and `purifying' the state there. Since the outer part of the boundary \TDH of the trapped region is time-like, there is no causal obstruction for these modes to continuously exit the trapped region across \TDH throughout the long evaporation process; the standard causal obstructions associated with EHs do not apply. This fact has been used in the literature to argue that there is no information loss issue at all \cite{hayward-pop,hayward-conf}; purification could have occurred all along the evaporation process. But this seemingly easy explanation is flawed. Examination of the renormalized energy flux shows that throughout this process, there is only infall across $\TDH$ in semi-classical gravity. Thus, the lack of a causal obstruction for the partner modes to exit the trapped region is not sufficient for the purification to occur in the semi-classical space-time.

In fact there is an apparent puzzle associated with the issue of information loss that is already relevant in the semi-classical regime. Since $M_{\rm moon} \sim  10^{-7}\, M_\odot$, at the end of this long evaporation process most of the initial ADM mass is carried away to $\scrip$ by the Hawking quanta. A back of the envelope calculation shows that a very large number $\mathcal{N}$\, ($\sim 10^{75}$)\, of quanta escape to $\scrip$ and all of them are correlated with the ones that fell across \TDH.  Therefore, at the end of the semi-classical process under consideration, one would have to have a huge number $\mathcal{N}$ of quanta both at $\scrip$ and in the trapped region, but the mass associated with the trapped region is only $10^{-7}$ times that carried away by the $N$ quanta going out to $\scrip$. Furthermore, the radius of the outer part of \TDH has shrunk to only $0.1$mm -- the Schwarzschild radius of a lunar mass black hole. How can a \TDH with just a $0.1$mm radius accommodate all these $\mathcal{N}$ quote? Even if we allowed each mode to have the (apparently maximum) wavelength of $0.1$mm, heuristically one would need the horizon to have a huge mass --some $10^{22}$ times the lunar mass!  While these considerations are quite heuristic, one needs to face the conceptual tension: At the end of the process under consideration, the trapped region has simply too many quanta to accommodate, with a tiny energy budget. Such considerations have led to suggestions that somehow `purification' must begin already by Page time  \cite{page} when the \TDH has lost only half its original mass of $M_{\odot}$, and essentially completed by the time the \TDH has shrunk down to the lunar mass. But this would imply that semi-classical considerations must fail in apparently tame regimes due to unforeseen quantum gravity effects that are relevant outside the horizons of astrophysical black holes! As we discussed in previous sections, in LQG quantum gravity corrections are completely negligible near horizons of macroscopic black holes.

The way out of the apparent paradox is that semi-classical theory itself predicts that the geometry of the trapped region has some rather extraordinary features that had not been noticed until relatively recently and not fully appreciated by the wider community even now. Calculations of the stress-energy tensor on the Schwarzschild space-times confirm the idea that, in semi-classical gravity there is a negative energy flux across the time-like portion of $\TDH$ such that $M_{\TDH}$ would decrease according to the standard Hawking formula: $\rmd M_{\TDH}/\rmd v = -  \hslash/(GM_{\TDH})^{2}$. (Indeed, this has been the basis of the standard estimate that the evaporation time goes as $\sim M^{3}_{\rm ADM}$.) One can then argue that, in the phase of evaporation from the solar mass to the lunar mass,  the form of the space-time metric in the trapped region of Fig.~\ref{fig:semi} is well approximated by the Vaidya metric:
\be \label{vaidya} {\rmd} s^{2}  = - \big(1- \f{2m(v)}{r} \big){\rmd} v^{2} + 2 {\rmd} v {\rmd} r + r^{2} \big({\rmd}\theta^{2} + \sin^{2}\theta\, {\rmd} \varphi^{2}\big), \ee
with $m(v) = GM_{\TDH}(v)$ decreasing \emph{very} slowly. As we saw in Section \ref{s2}, corrections due to quantum geometry are completely negligible in Schwarzschild interior until one reaches Planck curvature and, as Fig.~\ref{fig:semi} shows, that region is excluded in the semi-classical space-time under consideration. Thus, in the metric (\ref{vaidya}) quantum corrections are all induced by quantum matter and encoded in $m(v)$. To probe the geometry of the trapped region, it is convenient to foliate it and two natural foliations have been used by the LQG community: One defined by constancy of the Kretchmann scalar and the other by constancy of the radius of metric 2-spheres \cite{ao,aa-ilqg}. Each space-like slice is topologically $\mathbb{S}^{2} \times \mathbb{R}$ and is itself foliated by round 2-spheres which can be labelled by values of the advanced time coordinate $v$. Let us set $v=0$ when $M_{\TDH} = 1M_{\odot}$ and $v=v_{0}$ when $M_{\TDH}= M_{\rm moon}$.  During this long process, the radius of the MTSs on the outer, time-like part of \ATDH decreases from $r|_{v=0} = 3$km to $r|_{v=v_{0}} = 0.1$mm. The surprising fact is that as $v$ increases the leaves develop longer and longer necks of length $\ell_{N}$ along the $\mathbb{R}$ directions \cite{ao,cdl,mccr}. The `final leaf' for the process under consideration starts at the right end with $v=v_{0}$. The length $\ell_{N}$ of this final leaf is \emph{astonishingly large:}\, $\ell_{N} \approx 10^{64}$ \emph{light years} for the first foliation and $\ell_{N} \approx 10^{62}$ \emph{light years} for the second! These astronomically large lengths can result because the time the process takes is huge; $10^{64}$years corresponds to  $\sim\,10^{53}$ times our cosmic history!

This enormous stretching is analogous to expansion in (an anisotropic) cosmology. Recall that during the cosmic expansion --e.g. during inflation-- the wavelengths of modes get stretched enormously. This suggests that partner modes that fall into the trapped region will also get enormously stretched during evolution from $v=0$ to $v=v_{0}$, as in quantum field theory on an  expanding cosmological space-time, and become infrared. Can this phenomenon resolve the quandary of `so many quanta with so little energy'? The answer is in the affirmative. With such infrared wavelengths, it is easy to accommodate them in the trapped region with the energy budget only of $M_{\rm moon}$. Thus, even though the  outgoing modes carry away almost all of the initial mass  $M_{\odot}$ to $\scrip$, there is no obstruction to housing all their partners in the trapped region on a slice $\Sigma$ of Fig.~\ref{fig:semi} with the small energy budget of just $10^{-7}M_{\odot}$.  This argument removes the necessity of starting purification by Page time. In the LQG perspective, purification can be postponed to a much later stage.

To summarize, in the semi-classical regime, there are apparent paradoxes associated with the process of `purification' that is necessary for dynamics to be described by a unitary process. These disappear when one shifts the focus from event horizons to trapping dynamical horizons \emph{and} takes into account the time evolving geometry of the trapped region. By and large the LQG community has adopted this view.\medskip

\emph{Beyond the semi-classical regime:} When do quantum geometry effects become significant making the semi-classical approximation inadequate? The viewpoint in LQG is that this happens when physically observable quantities such as curvature scalars and matter density enter the Planck regime. This expectation was borne out in the investigation of the Schwarzschild interior in Section \ref{s2}. Therefore, one would expect semi-classical considerations to be valid well beyond the time when \TDH has shrunk to $M_{\TDH} = M_{\rm moon}$ we considered in the above discussion, all the way till the curvature is, say, $10^{-6}$ times the Planck curvature which corresponds to $M_{\TDH} \approx 10^{3} M_{\rm Pl}$. LQG explorations of the evaporation process beyond this stage are being carried out by different groups. The main ingredients are: results on causal structure of the Schwarzschild interior summarized in Section \ref{s2}, intuition derived from simpler models such as CGHS \cite{atv}, conclusions drawn from a long series of works (see, e.g., \cite{hayward,frolov,bardeen,ebms,crfv,ebtdlms,hhcr,mc,bcdhr,mdcr}) that posit a space-time structure for the entire process and work out its consequences, strong consistency requirements on the ensuing space-time geometry (see, e.g., \cite{Carballo-Rubio:2019nel}), and calculations based on the Vaidya metric for the structure of space-time in the \emph{distant} future \cite{ori4,ebtdlms}. While there is broad consensus on the overall picture, many open issues remain. We will now summarize the current status.

\begin{figure}[b]
%\sidecaption[t]
\includegraphics[width=0.3\textwidth]{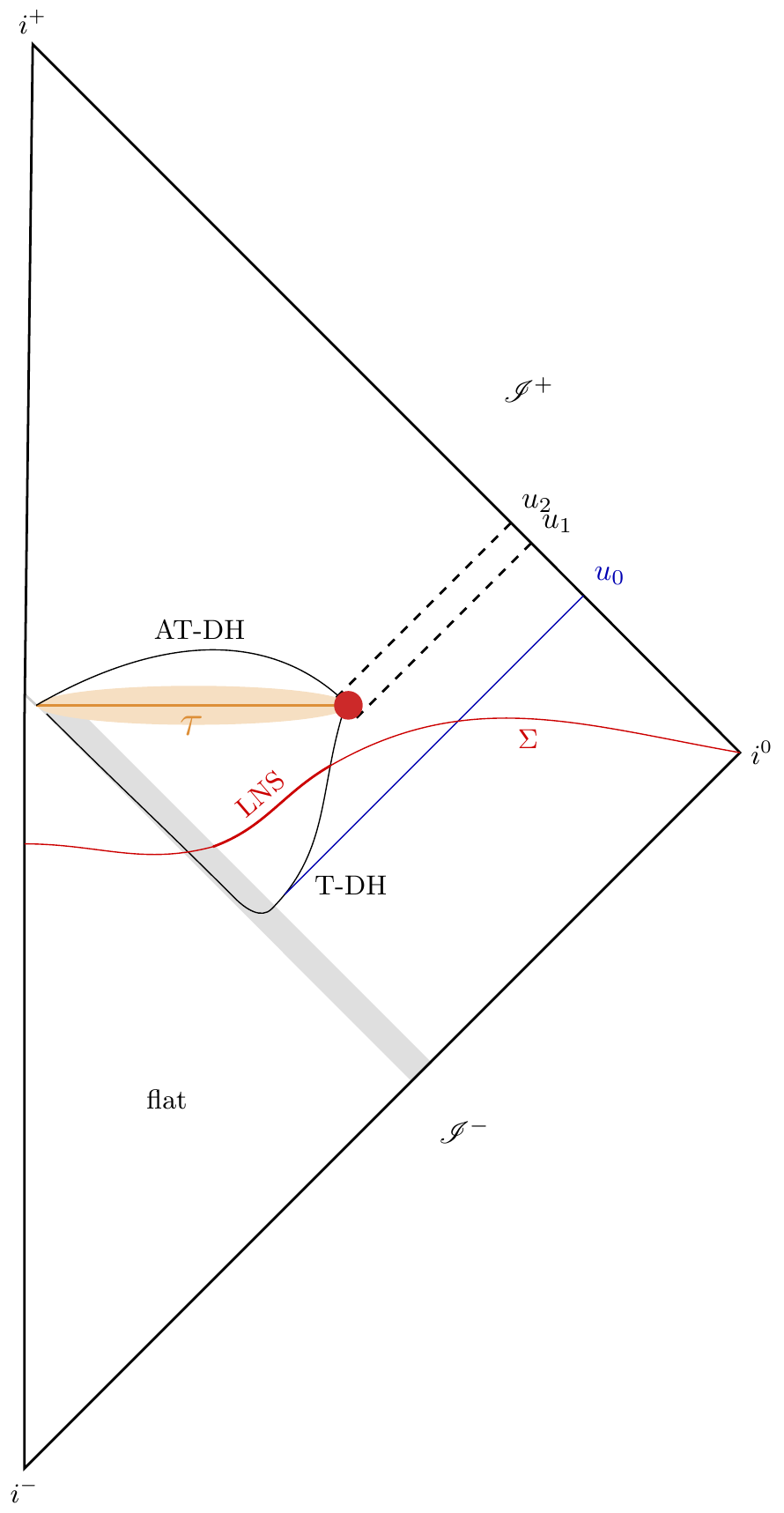}
%
%\caption{\footnotesize
\caption{\footnotesize{Quantum extension of the space-time in LQG.  The classical singularity is replaced by a \emph{Transition surface} $\tau$, to the past of which we have a trapped region, bounded in the past by a trapping dynamical horizon \emph{T-DH}, and to the future of which we have an anti-trapped region bounded by an anti-trapping dynamical horizon \emph{AT-DH}. Cauchy surfaces $\Sigma$ develop astronomically long necks already in the semi-classical region. The dark (red) blob at the right end of $\tau$ is a genuinely quantum region. }}
\label{fig:LQG}
\end{figure}

To the future of the semi-classical region, curvature can exceed $10^{-6} \lp^{-2}$, whence we need full quantum gravity. This region with Planck scale curvature is depicted by the shaded (pink) region in Fig.~\ref{fig:LQG}. (To the past and future of this region, semi-classical gravity should yield a reasonable approximation.)  In the shaded (pink) region geometry is described by a quantum state $\Psi_{\rm geo}$ and the difficult task is to evolve the quantum field $\hat\phi$ on this quantum geometry. Fortunately, prior experience with other systems --such as the propagation cosmological perturbations on the quantum FLRW geometry-- suggests a strategy that is applicable during the adiabatic phase of the Planck regime. The evaporation process is adiabatic so long as mass-loss does not occur too rapidly, i.e., until the radius of \TDH is $\approx 10^{3} \lp$. At the end of this process, one enters a neighborhood of the future endpoint of the \TDH depicted by a (red) blob, where the curvature is Planckian \emph{and} the process speeds up very rapidly, violating the adiabatic approximation. Let us first discuss the adiabatic phase and then return to the (red) blob.
\nopagebreak[3]

Prior results in LQC strongly suggest that the problem of propagating a quantum field $\hat\phi$ on a quantum geometry represented by $\Psi_{\rm geo}$ can be greatly simplified during the adiabatic phase: One can construct a smooth $\t{g}_{ab}$ that carries all the information in $\Psi_{\rm geo}$ that the dynamics of quantum fields $\hat\phi$ is sensitive to (see, e.g., \cite{aan3}). $\t{g}_{ab}$ is called the \emph{dressed metric}. Thus, the difficult task of evolving quantum fields on quantum geometry is reduced to that of evolving them on the space-time of the dressed metric $\t{g}_{ab}$. Next, the expectation from results of Section \ref{s2} is that the shaded (pink) region will contain a transition surface $\T$  (w.r.t. $\t{g}_{ab}$) that replaces the classical singularity and separates the trapped region that lies to its past and the untapped region that lies to its future. The metric $\t{g}_{ab}$ will capture two distinct effects: those that originate from quantum geometry and feature the area gap $\Delta$ (as in Section \ref{s2}), and those that are induced on $\t{g}_{ab}$ by the falling quantum matter, dominated by the incident pulse of the scalar field at the left end of the (pink) shaded region, and by the infalling Hawking quanta carrying negative energy as one moves towards the right end.

Discussion of Section \ref{s2} strongly suggest that the first set of effects will decay rapidly as we move away from Planck curvature into the semi-classical region. Therefore in the semi-classical region, $\t{g}_{ab}$ will be well approximated by $g^{\rm sc}_{ab}$ used there. As we move to the future of the (pink) shaded region, one would encounter an anti-trapping dynamical horizon \ATDH (see Fig.~\ref{fig:LQG}). The region enclosed by the transition surface $\T$ to the past and \ATDH to the future would be anti-trapped as in Fig.~\ref{fig:diamond}. But now \ATDH  would be space-like rather than null and its area would not be constant, but decrease as one moves left. Qualitatively this change in the structure of \ATDH from the one of Fig.~\ref{fig:diamond} is parallel to the change in the structure of the trapping horizon \TDH that we already discussed in some detail.

Finally, the region to the future of \ATDH would also be well approximated by a Vaidya metric, but now the outgoing one, expressed in terms of the retarded time coordinate $u$ in place of the advanced time coordinate $v$ of Eq. (\ref{vaidya}). It will describe the propagation of the infrared modes that will emerge from the \ATDH and arrive at $\scrip$ at very late times. (The metric in this region will be nearly flat because the total energy in the scalar field is small and dispersed over very large spatial regions.) Recall that these are the partner modes that fell into the horizon and were therefore entangled with the outgoing modes that carried away most of the initial ADM mass. In the LQG scenario, then, correlations are finally restored at $\scrip$ where, in the end, the infalling modes also arrive. The total energy carried by the two sets of modes is very different. But this is not an obstruction for restoring correlations, i.e., for the `purification' to occur. The timescale of this purification process is very long, ${0}(M^{4})$ \cite{ori4,ebtdlms,bcdhr}. Purification can occur much later than the page time because of the LQG singularity resolution.

The final picture is rather similar to the process of burning a piece of coal that is often invoked in the discussion of black hole evaporation. Initially the piece of coal is in a pure state. When lit, it emits photons  and the energy they carry is well described by a thermal state at high temperature. But the total state is pure because there are correlations between the quantum state of outgoing photons and the left-over coal. As the fire extinguishes, there is very little energy left and the ashes emits photons with lower and lower frequencies for a very long time. At the end of the process the cold ashes are in a pure state and all photons have escaped. The late time, long wavelength photons are able to restore the correlations that were apparently lost in the middle of the process (when the photon spectrum seemed approximately thermal) even though the total energy they carry is small compared to the energy carried by high frequency photons that were emitted earlier.

Note, however, that this scenario is incomplete because one still has to deal with the very last part of the evaporation process, depicted by the red blob and the associated null rays $u=u_1$ and $u =u_2$ of Fig. \ref{fig:LQG}. In this region,  not only is the curvature of Planck scale, but it is varying extremely rapidly because it lies at the end point of the evaporation process. It is this combination that makes the problem difficult; if we had only one of these features, we could have used known approximation methods. Independent considerations suggest that something very non-trivial must happen in this region. We will conclude with an example. During the semi-classical phase, as the \TDH shrinks, the temperature associated with the radiation at $\scrip$ grows. Consequently, the modes become increasingly ultraviolet as one approaches the point $u=u_1$ on $\scrip$. On the other hand, the radiation that emerges from the anti-trapped region is infrared and received at $\scrip$ to the future of $u=u_2$. This is a dramatic transition, strongly suggesting that the physics of the region which is highly dynamical \emph{and} has Planck scale curvature will be very subtle and interesting. %It remains an attractive challenge in LQG.
For example, it has been suggested that Planck scale `seeds' may be left behind, scattered in this region \cite{perez}. Understanding the nature of this quantum geometry remains an attractive challenge in LQG.

Let us summarize the current status of LQG investigations of black-evaporation. They are distinguished by their emphasis on two features that are generally ignored in other approaches: (i) A shift away from the teleological event horizons \emph{EH}  to quasi-locally defined trapping and anti-trapping DHs \TDH and \ATDH\!;\, and,\, (ii) replacement of the classical singularity by the transition surface $\T$. As a consequence, the traditionally used Penrose diagram of Fig.~\ref{fig:trad}\, is replaced by the Penrose diagram of Fig.~\ref{fig:LQG}.

\section{Discussion}
\label{s6}

As the bibliography indicates, LQG literature on regular black holes is very rich. Indeed, even this long list is far from being exhaustive! To make the material accessible to non-experts, we focused on four lines along which advances have occurred, and in each case built the discussion around a few of the mainstream developments. Much of this  discussion is based effective equations, motivated by the fact that high performance computations have shown that effective space-time metrics provide an excellent approximations to the quantum geometry in LQC, also in Bianchi models where the Weyl curvature is non-zero and diverges at the singularity \cite{Diener:2014mia, Diener:2017lde}.

The first area, discussed in Section \ref{s2}, focuses on the `Schwarzschild interior' that contains the singularity. Since the resolution of this singularity is central to the theme of `regular black holes' of this Volume, we included  an account of several different effective descriptions. These investigations bring out two features: (i) singularity resolution due to the underlying quantum geometry effects of LQG is robust, and does not depend on details of the quantization methods;\, but,\, (ii) the precise manner in which quantization is carried out can unleash unintended and physically undesirable effects that are not apparent until a detailed examination is carried out. We summarized a scheme that is free of these drawbacks. The resulting quantum corrected geometry exhibits interesting causal structures: the singularity is replaced by a transition surface,\, $\T$,\, to the past of which there is a trapped region, and to the future, an anti-trapped region. Each is bounded by null horizons and, for macroscopic black holes, the area of the future horizon is approximately equal to that of the past (see Fig.~\ref{fig:diamond}). In each of these effective geometries, curvature scalars attain their maxima at the transition surface which, furthermore, have \emph{universal} values, independent of the mass of the black hole. This universality seems to be a general feature of the singularity resolution due to quantum geometry effects of LQG.

Section \ref{s3} extended the quantum corrected geometry of Section \ref{s2} to the exterior, asymptotic region of the Schwarzschild space-time by exploring the homogeneity of time-like surfaces $r_{\rm sch} ={\rm const}$. For macroscopic black holes (i.e., those with $m \gg \lp$), the near horizon geometry of this exterior has the expected and physically desired features: the quantum corrected, effective metric is smooth across the horizon and corrections to the Hawking temperature, computed using methods from Euclidean quantum field theory, are tiny. More generally, for macroscopic black holes there is excellent agreement between the effective geometry and that of the classical Schwarzschild metric in a vast neighborhood of the horizon in the exterior region. Unfortunately, there is some confusion in the literature on this point arising from the simplified form (\ref{tg-asym}) of the metric that holds in the far-asymptotic region, i.e., \emph{only on ignoring terms ${O}(r_s/r)$}. If one overlooks this key approximation and uses (\ref{tg-asym}) in the entire exterior region --as was done in \cite{faraoni}-- one obtains ``unsettling features", such as non-trivial corrections to the innermost circular orbit. These are consequences not of the actual effective geometry, but of the incorrect use of its simplified form. (Nonetheless, unfortunately, incorrect conclusions of \cite{faraoni} have been repeated in some of the subsequent literature, e.g. \cite{bojowald:2020}). More generally, the near horizon quantum corrections to astrophysical black holes will be very small to have observable relevance in the foreseeable future (at least in the non-rotating case on which most LQG investigations have focused so far).%
\footnote{In this review, we did not touch on the issue of black hole entropy that arises in LQG by counting microstates of the area operator that are compatible with parameters characterizing a given macroscopic black hole (see. e.g., \cite{perez-review}. The possibility of testing discreteness of area using gravitational waves has drawn considerable attention in the literature. It has been argued that the simplest area spectrum with area eigenvalues given by $k n\, \lp^2$  (where $n$ is an integer and $k$ a constant), considered by Bekenstein and Mukhanov \cite{jbvm}, could be ruled out using data from a sufficiently large number of compact binary mergers. But in LQG the area spectrum is \emph{not} equidistant, it crowds exponentially, making the continuum an excellent approximation very quickly. However, for small black holes the area eigenvalues are grouped, exhibiting a band structure, and the separation between bands is ${O}(\lp^2)$. If this structure were to persist for large rotating black holes, each band would serve as a proxy of the Bekenstein-Mukhanov eigenvalues and gravitational observations would then lead to non-trivial constraints \cite{acdrmp}. However, currently there is no evidence that points to the persistence of bands for macroscopic areas.}

The full metric in the exterior region is also asymptotically flat with curvature decay that is sufficient for the ADM mass to be well defined (e.g., if one uses the expression (\ref{ERicci}) in terms of the spatial Ricci tensor). For macroscopic black holes, quantum corrections to the classical value are very small. However, the decay is slower than that in the standard notion of asymptotic flatness. Consequently, different expressions of the ADM mass, that must agree with one another exactly if the standard asymptotic conditions hold \cite{aaam2}, now differ by quantum corrections. Much more surprising is the feature that the norm of the time-translation Killing field of the effective metric diverges at spatial infinity! One's first reaction would be that such deviations from standard asymptotic flatness must lead to a plethora of physically inadmissible consequences. One test is provided by quasi-normal modes. Do they exhibit a pathological behavior? A detailed investigation \cite{kunstatter} has shown that the potential which enters the quasi-normal mode analysis continues to be well-defined everywhere. One can then compute quasi-normal frequencies using an approximation tailored to improving accuracy. The corrections to the classical result are found to be negligibly small. An independent investigation \cite{dco} provided  expressions for axial and polar perturbations, computed their quasi-normal frequencies and found departures with respect to the classical theory; in particular, isospectrality is broken. However, all these relative deviations from the classical predictions are only a small-percent effect even for black holes as small as $r_S\sim 10^3 \lp $, and they decrease with the mass of the black hole, becoming completely negligible for macroscopic black holes. These investigations also show that the metric passes the stability criterion for tensor and massless scalar field perturbations. However, the infrared behavior of the potential is different from that in the classical Schwarzschild case, leading to a qualitative difference in the power-law tails. These tails play an important role in the mathematical literature but are not astrophysical significant because they occur after the waves are exponentially damped in the quasi-normal ringing phase. In summary, at present it is not clear whether counter-intuitive features associated with the asymptotic behavior of the effective metric of \cite{aos} are indications that it may  be inadmissible in the asymptotic region $r_S/r \ll 1$, or if they are physically harmless. In view of this uncertainty, several investigations are exploring alternate ways of arriving at an effective metric that has the standard asymptotic behavior (see, e.g., \cite{gop:2021,han}).

Another conceptual issue concerns covariance. There \emph{is} a 4-metric in the full quantum extended Kruskal space-time and results on singularity resolution, for example, refer to curvature invariants; these considerations are all 4-dimensionally covariant. But to arrive at the effective Einstein's equations with quantum geometry modifications, one uses symmetry reduction. The question is whether there is a covariant action for the full theory \emph{without symmetry reduction} whose equations of motion reduce to those in Sections \ref{s2} and \ref{s3} in its static, spherically symmetric sector \cite{bojowald:2019}. This is a technically difficult issue that is still open. Indeed, it took some time to show that the much simpler effective equations of the homogeneous, isotropic sector of LQC \cite{aos} can be obtained in this way, but finally the answer turned out to be in the affirmative \cite{olmo-singh}. A similar situation arose also in string theory where it seemed for quite some time that the exact 1+1 dimensional stringy black hole \cite{dvv} did not arise from the symmetry reduction of a covariant action \cite{grumiller:2002}. In the end, it was shown that there is such an action but it requires inclusion of additional fields \cite{grumiller:2005}. There are some concrete indications that the situation is likely to be similar in the LQG black hole sector we discussed; see e.g., \cite{han} that introduces a covariant action in the `mimetic gravity' setting that adds a scalar field with a specific potential and uses the same time-like homogeneous slices as in Section \ref{s3} in the symmetry reduced sector, and \cite{Alonso-Bardaji:2021yls} that uses a reasoning based on the constraint algebra to argue for covariance.

Discussion in Sections \ref{s2} and \ref{s3} was confined to the LQG treatment of the eternal black hole and arrived at the Penrose diagram of Fig.~\ref{fig:penrose} for the quantum extension of the Kruskal space-time. This entire space-time is non-singular. In classical relativity as well as in the discussion of black hole evaporation, Kruskal space-time of Fig.~\ref{fig:kruskal} provides useful mathematical tools as well physical intuition. The same is true  of its quantum extension. However, realistic and more interesting situations involve formation of black holes by gravitational collapse (for which only a part of the full Kruskal space-time is relevant). In Section \ref{s4} we focused on two complementary issues that have been investigated in dynamical situations featuring gravitational collapse. The first involves the resolution of singularity for collapsing dust models. Here, the emphasis is on quantum geometry effects because the matter is characterized by dust rather than a fundamental quantum field. Thus, non-classical features associated with matter --such as boundedness of the dust density-- are induced on matter by the quantum nature of geometry. These investigations show that, as in cosmology, the singularity is replaced by a bounce; this is a robust result. The second class of investigations focuses on critical phenomena. Now the strategy is the opposite in that it is matter that is represented by a quantum scalar field of LQG \cite{Ashtekar:2002vh} while geometry is classical to begin with; corrections to classical effects on geometry are induced by quantum matter through field equations.
The overall finding is that the quantum corrections to the classical results are small for macroscopic black holes, just as one would hope. However, while there is no `mass-gap' in the classical theory\, --i.e. a black hole can be formed with arbitrarily small mass--\, a mass gap can develop if one uses a quantization scheme that leads to effective equations violating scale invariance.

While dynamics is at the heart of these investigations, they do not encompass the Hawking process because, in the first set of analyses matter is classical, and the second focuses on critical behavior in gravitational collapse, rather than on the scalar field quanta going out to $\scrip$, or the issue of entanglement. LQG investigations of the evaporation process\, --including the issue of back reaction on geometry--\, were discussed in Section \ref{s5}. They reflect a broad consensus that the arguments that lead to the traditional Penrose diagram of Fig.~\ref{fig:trad} are flawed in two important respects. First, they assume that a part of classical singularity persists in the quantum theory while it is resolved in LQG.

Second, the event horizon plays a key role in Fig.~\ref{fig:trad} even though it is teleological and can be made to disappear by changing space-time geometry in a Planck scale neighborhood of the singularity \cite{ph}. The traditional Penrose diagram is replaced by a new LQG Penrose diagram shown in Fig.~\ref{fig:LQG}. There is consensus that there is no information loss: The S-matrix from $\scrim$ to $\scrip$ is unitary provided, of course, we consider a closed system in which the black hole forms by the gravitational collapse of a quantum field from $\scrim$ and we use the quantum state of the same field at $\scrip$. That the singularity would be resolved by quantum geometry effects is motivated by two considerations: (i) Quantum geometry effects discussed in Section \ref{s2} that provide universal upper bounds to curvature scalars because of a non-zero value of the area gap; and, (ii) detailed numerical simulations in the CGHS case that show that even in the semi-classical theory, the singularity is significantly weakened when back reaction effects are included, which already suffice to make the metric continuous there \cite{apr}. There is no\, \emph{EH}\, in the final picture; \emph{what forms classically in the gravitational collapse and evaporates through quantum processes is a}\,\emph{DH}.

The evaporation process of LQG can be described as follows. The Hawking quanta are created in pairs, the outgoing quanta go out to $\scrip$ as in Hawking's original paradigm, and their partner quanta fall across the trapping dynamical horizon \emph{T-DH}. In the semi-classical regime depicted in Fig.~\ref{fig:semi}, the outgoing quantum state is well approximated by a thermal state at $\scrip$ (at sufficiently late times), and the partner modes carry a negative energy flux into the trapped region that is bounded by \TDH in this figure. When the back reaction is included, the geometry in the trapped region changes adiabatically, and space-like surfaces $\Sigma$ of Fig.~\ref{fig:LQG} get stretched and become long necked surfaces (\emph{LNS}). Suppose that at its formation, the black hole has solar mass $M_\odot$. Although the process of elongation of necks is very slow, it continues for a very long time since the semi-classical phase lasts some $10^{64}$ years. At the end of this phase, the necks become astronomically long, stretched to some $10^{62}$ light years! Therefore the modes that have fallen in the trapped region also get enormously stretched (as they do during inflation) and become infrared. They can continue to be entangled with their partner modes that went out to $\scrip$ during this long semi-classical phase\, --even though the total energy carried by the outgoing modes is almost $M_\odot$ and that carried by the trapped modes is tiny--\, precisely because the trapped modes are infrared. Thus, the quantum state on a surface such as $\Sigma$ continues to be pure. Since the singularity is resolved and replaced by a transition surface $\T$ that lies in the shaded (pink) region, these modes can evolve across the transition surface $\T$ and emerge on the other side. Then they propagate to the approximately flat region that lies to the future of the anti-trapped dynamical horizon $\ATDH$, and arrive at $\scrip$ restoring the correlations with the partner modes that reached $\scrip$ much earlier, during the semi-classical phase. (As explained towards the end of section \ref{s5.2}, this situation is qualitatively similar to that of burning a piece of coal where correlations are restored at late times when the large wavelength modes emerge from ashes as they cool down, restring correlations with short wavelength modes emitted earlier.)

What remains largely unexplored so far is the (red) `blob' at the right end to the shaded (pink) region in Fig. \ref{fig:LQG} and how it affects the physics at $\scrip$. As discussed at the end of Section \ref{s5}, the problem is hard because one simultaneously encounters two difficulties: Planck scale curvature and rapid changes that make adiabatic approximation inadequate. But feasible calculations may suffice to reveal whether most, if not all, of the correlations are restored when the infrared modes traverse the anti-trapped region and emerge at $\scrip$. If they are restored, then the fully quantum `blob' would not be that relevant for the issue of information loss and the S-matrix would be unitary. Although the consensus in LQG favors this possibility, this issue is open. There \emph{are}  arguments involving a fully quantum evolution from past of the blob to its future, but so far they are inconclusive because of the underlying assumptions. At this stage, one cannot, for example, rule out the possibility that the `blob' joins on to a baby universe whose states are inaccessible from $\scrip$ of Fig.~\ref{fig:LQG}. If this were ti happen,  from the perspective of $\scri^\pm$ of this figure, information may be lost, although the `total' S-matrix would be unitary. Showing that this does not happen remains a fascinating challenge in the LQG community.\medskip

Let us summarize. The LQG community has explored different aspects of the many fascinating properties of black holes. The distinguishing feature of these investigations is their emphasis on quantum geometry that is directly responsible for replacement of the singularity by a transition surface with interesting causal properties, and  boundedness of physical observables such as curvature scalars and matter density. As discussed in Section \ref{s1}, these features are not shared by other approaches: Using the AdS/CFT correspondence as motivation, it is sometimes argued that singularities should persist also in quantum gravity, and indeed, much of the literature uses the Penrose diagram of Fig.~\ref{fig:trad} in which a singularity features as part of the future boundary of space-time.
On the other hand, because quantum geometry effects become important only in the Planck regime, LQG corrections to the classical results are very small near the horizons of astrophysical black holes; examples we discussed include corrections to the Hawking temperature using the near horizon geometry and the machinery of Euclidean quantum field theory, corrections to quasi-normal frequencies of astrophysical black holes, and to results associated with critical collapse. This is also in striking contrast to some other approaches, e.g., the `firewall scenario' that emerged from string theory  considerations. More generally, LQG does not lead to violations of semi-classical expectations of physics near horizons of astrophysical black holes that had been advocated before the LIGO discoveries showed that predictions of classical GR, without such major corrections, are realized in compact binary mergers. As mentioned in Section \ref{s5.2}, there is also a large body of investigations that posit a space-time structure for the entire process and work out its consequences. By and large one solves \emph{classical} Einstein's equations (with suitable stress-energy tensors) in various patches, and joins them consistently. Some of these space-time diagrams resemble Fig.~\ref{fig:LQG}. While these investigations do pay close attention to consistency conditions, and often also to energy considerations, the issue of \emph{quantum correlations and unitarity} received little attention in these works. The LQG line of reasoning of Section \ref{s5} fills this conceptually important gap that is key to the issue of `information loss'.

There is also considerable discussion on the issue of young versus old black holes, and long lived remnants. In LQG, there is indeed an important difference between a young and an old black hole. As a concrete example, let us consider two lunar mass black holes -- a young one that is freshly formed from gravitational collapse, and an old one what started out as a solar mass black hole and then evaporated down to the lunar mass, as discussed in Section \ref{s5.2}. While their dynamical horizons will have the same radius, $0.1$mm, and mass $M_{\TDH} = M_{\rm moon}$, their  external environment  as well as internal structure will be \emph{very} different. In the second case, the evaporation process would have gone on for some $10^{64}$ years. Therefore, there will be a very large number of outgoing Hawking quanta in the exterior region, and an equal number of ingoing quanta in the trapped region, the two being entangled. Therefore, the small area of \TDH will not be a measure of the entropy of what is in the interior (or exterior). However, in LQG, in both cases the area is a measure of the surface degrees of freedom of the horizon, i.e., degrees of freedom that can communicate both the outside and inside regions. But it is sometimes argued that there is a potential problem with this scenario: because old black holes can have small energy but an enormous number of modes, it should be easy to produce them in particle accelerators. But these arguments use only the conservation laws normally used in computing scattering amplitudes; since old black holes have astronomically long necks, it is hard to imagine how such changes in space-time structure can occur on time scales of accelerator physics \cite{ori1}.

Finally, let us discuss some of the limitations of the current LQG investigations. The analyses we summarized make a strong use of symmetry reduced models and effective equations that capture the leading order quantum corrections. However, there have been a number of interesting investigations that aim at arriving at these effective equations starting from full LQG (see, e.g., \cite{alescietal,Assanioussi:2019twp}). But they are still in a rather preliminary stage, and further and more detailed investigations are needed. Another key limitation is that so far the LQG investigations have focused primarily on non-rotating black holes, where the classical singularity is space-like. But for rotating black holes the inner horizons would be unstable and therefore the singularity would be null. So far quantum geometry considerations have not been applied to null singularities. This is an outstanding open problem. Indeed, inclusion of rotating black holes in the discussion of `information loss' during  evaporation remains a fascinating problem in \emph{all} approaches to quantum gravity.

\section*{Acknowledgements}
%\begin{acknowledgement} 
This work was supported in part by the NSF grant PHY-1806356, PHY-1912274 and PHY-2110207, Penn State research funds associated with the Eberly Chair and Atherton professorship, and by Projects PID2020-118159GB-C43, PID2019-105943GB-I00 (with FEDER contribution), by the Spanish Government, and also by the ``Operative Program FEDER2014-2020 Junta de Andaluc\'ia-Consejer\'ia de Econom\'ia y Conocimiento'' under project E-FQM-262-UGR18 by Universidad de Granada. We would like to thank Eugenio Bianchi, Kristina Giesel, Muxin Han, Bao-Fei Li, Guillermo Mena, Sahil Saini and Ed Wilson-Ewing for discussions, and Tommaso De Lorenzo for Figures 4-6.
%\end{acknowledgement}

%%%%%%%%%%%%%%%%%%%%%%%%%%%%%%%%%%%%%%%%%%%%%%%%%%%%%%%%%

\end{document}